\documentclass [superscriptaddress,floatfix,showpacs,preprint] {revtex4}

\usepackage {amsmath,amssymb,epsfig,color}

\begin{document}

\title
{Full orbital calculation scheme for materials with strongly correlated
electrons}

\author{V.I.~Anisimov}
\affiliation{Institute of Metal Physics, Russian Academy of
Sciences-Ural Division, 620219 Yekaterinburg GSP-170, Russia}
\author{D.E.~Kondakov}
\affiliation{Institute of Metal Physics, Russian Academy of
Sciences-Ural Division, 620219 Yekaterinburg GSP-170, Russia}
\author{A.V.~Kozhevnikov}
\affiliation{Institute of Metal Physics, Russian Academy of
Sciences-Ural Division, 620219 Yekaterinburg GSP-170, Russia}
\author{I.A.~Nekrasov}
\affiliation{Institute of Metal Physics, Russian Academy of
Sciences-Ural Division, 620219 Yekaterinburg GSP-170, Russia}
\author{Z.V.~Pchelkina}
\affiliation{Institute of Metal Physics, Russian Academy of
Sciences-Ural Division, 620219 Yekaterinburg GSP-170, Russia}

\author{J.W.~Allen}
\affiliation{Randall Laboratory of Physics, University of
Michigan, Ann Arbor, MI 48109}
\author{S.-K.~Mo}
\affiliation{Randall Laboratory of Physics, University of
Michigan, Ann Arbor, MI 48109}

\author{H.-D.~Kim}
\affiliation{Pohang Accelerator Laboratory, Pohang 790-784, Korea}

\author{P.~Metcalf}
\affiliation{Department of Physics, Purdue University, West Lafayette, IN 47907 USA}

\author{S.~Suga}

\affiliation{Division of Materials, Physics Graduate School of
Engineering Science Osaka University Toyonaka, Osaka 560-8531
Japan}
\author{A.~Sekiyama}
\affiliation{Graduate School of
Engineering Science, Osaka University Toyonaka, Osaka 560-8531
Japan}
\author{G.~Keller}
\affiliation{Theoretical Physics III, Center for Electronic
Correlations and Magnetism, University of Augsburg, 86135
Augsburg, Germany}
\author{I.~Leonov}
\affiliation{Theoretical Physics III, Center for Electronic
Correlations and Magnetism, University of Augsburg, 86135
Augsburg, Germany}
\author{X.~Ren}
\affiliation{Theoretical Physics III, Center for Electronic
Correlations and Magnetism, University of Augsburg, 86135
Augsburg, Germany}
\author{D.~Vollhardt}
\affiliation{Theoretical Physics III, Center for Electronic
Correlations and Magnetism, University of Augsburg, 86135
Augsburg, Germany}

\date{\today}

\begin{abstract}

We propose a computational scheme for the  {\it{ab initio}}
calculation of Wannier functions (WFs) for correlated electronic materials.
The full-orbital Hamiltonian $\hat H$ is projected
into the WF subspace defined by the physically most relevant
partially filled bands. The Hamiltonian $\hat H^{WF}$
obtained in this way, with interaction parameters
calculated by constrained LDA for the Wannier orbitals, is used as an {\it{ab initio}}
setup of the correlation problem, which can then be solved by many-body techniques,
e.g., dynamical mean-field theory (DMFT). In such calculations
the self-energy operator $\hat\Sigma(\varepsilon)$ is defined in
WF basis which then can be converted back into the
full-orbital Hilbert space to compute the
full-orbital interacting Green function
$G({\bf r},{\bf r}^\prime,\varepsilon)$.
Using $G({\bf r},{\bf r}^\prime,\varepsilon)$ one
can evaluate the charge density, modified by correlations, together
with a new set of WFs, thus defining a fully self-consistent scheme.
The Green function can also be used for
the calculation of spectral, magnetic and electronic properties of the
system. Here we report the results obtained with this method for SrVO$_3$ and
V$_2$O$_3$. Comparisons are made with previous results
obtained by the LDA+DMFT approach where the LDA DOS was used as input,
and with new bulk-sensitive
experimental spectra.

\end{abstract}
\maketitle

%%%%%%%%%%%%%%%%%%%%%%%%%%%%%%%%%%%%%%%%%%%%%%%%%%%%%%%%%%%%%%%%%%%%%%%%%%%%
% Introduction
%%%%%%%%%%%%%%%%%%%%%%%%%%%%%%%%%%%%%%%%%%%%%%%%%%%%%%%%%%%%%%%%%%%%%%%%%%%%
\section{Introduction}
\label{intro}

Model Hamiltonians used in the study of correlation effects in
solids have a Coulomb interaction term in a site-centered atomic-like
orbital basis set which is not explicitly defined. When the
correlated electrons are well localized, as, for example,
4$f$-states of rare-earth ions, atomic orbitals (or atomic
sphere solutions like muffin-tin orbitals) are a good choice.
However, the most interesting problems occur in the regime of
metal-insulator transitions, where the states of interest become
partially itinerant and rather extended. The error of using atomic
orbitals is most severe in the case of materials with strong
covalency effects, like late transition metal oxides, where
partially filled bands are formed by the mixture of metallic
$d$-orbitals and oxygen p-orbitals. For example, in high-T$_c$
cuprates correlated states have the symmetry of Cu-3$d$ $x^2-y^2$
orbitals, but are actually Zhang-Rice singlets formed by the
combination of oxygen p-states centered around the Cu ion and having
$x^2-y^2$ symmetry.

In model calculations the problem of defining the correlated orbitals
is not very important, because it only affects model
parameter values, which in any case are considered fitting
parameters. However any attempt to construct an
``{\it{ab initio}}'' calculation scheme requires an explicit
definition of the basis set for the Coulomb interaction term. An
important requirement for such a choice is that the orbitals must
produce the partially filled bands where Coulomb correlations
occur while preserving the localized, site-centered atomic-like form.
These requirements are fulfilled for Wannier functions (WFs)
$|W_{n}^{{\bf T}}\rangle$ defined as a Fourier transformation of
the Bloch functions $|\psi_{n\bf k}\rangle$ \cite{wannier}. Here
and below functions are labeled with band index n, lattice
translation vector {\bf T} and wave vector {\bf k}.

When there is more than one band crossing the Fermi level, WFs
are not uniquely defined. Any {\bf k}-dependent unitary
transformation $\hat U^{(\bf k)}$ of the set of Bloch functions
$|\psi_{n\bf k}\rangle$ for these bands produces a new set which
can be used for the calculation of WFs via Fourier
transformation (eq.~\ref{psi_def}, Sec.~\ref{W_def}). If one
imposes the requirement that the WFs should have the
symmetry of atomic orbitals~\cite{vanderbildt,pickett}, this
unitary transformation is well defined. The explicit form of
the WFs allows one to compute Coulomb interaction parameters
in constrained local density approximation (LDA)
calculations.

In this way the parameters for the {\it{ab initio}} many-body Hamiltonian
(non-interacting Hamiltonian $\hat H^{WF}$ and Coulomb interaction)
in the WF basis can be computed by any first principle
electronic structure calculation scheme (below we use the LMTO method).
This Hamiltonian can then be further investigated by one of the methods developed in
the many-body community. In the present work we use the dynamical
mean-field theory (DMFT)~\cite{vollha93,pruschke,georges96}. Within
DMFT, the effective impurity problem corresponding to the many-body Hamiltonian
is solved by quantum Monte-Carlo
simulations (QMC) ~\cite{QMC}. The DMFT part of the proposed calculation
scheme is essentially the same as the one used in the recently developed
LDA+DMFT approach~\cite{LDADMFT1}
for the {\it{ab initio}} investigations of correlated electron
materials~\cite{LDADMFT}. However, here we propose a more general procedure
to compute the Green
function using the Hamiltonian matrix and an integral over
Brillouin zone instead of the Hilbert transform of the LDA density
of states (DOS). This particular method allows one to avoid the uncontrollable
errors occuring in the computation of the Green function using the Hilbert
transform of the LDA DOS. Thus, to obtain an insulating solution we
need to cut off the long (metal-oxygen) hybridization tails of the DOS,
renormalize it and shift the Fermi energy to get an integer filling.
In the present method we overcome the above-mentioned difficulties owing to
the integer filling of Wannier orbitals. The result of the
DMFT calculations is a local self-energy operator $\hat
\Sigma(\varepsilon)$. In our scheme this operator is defined in the WF
basis set $\{W_n\}$ and is acting in the subspace of partially filled
bands which are used for the construction of the WFs.

The paper is structured as follows.  In Sec.~\ref{method} the
details of our scheme are presented. In Sec.~\ref{W_def} we
describe the construction of WFs, as well as the {\it ab-initio}
Hamiltonian matrix within this basis set in terms of Bloch
functions. In Sec.~\ref{WF_GF} we propose a general method for the
construction of WFs using the Green function $G(\bf r,\bf
r^\prime,\varepsilon)$, which reduces to the results of
Sec.~\ref{W_def} in the non-interacting case. The reason for doing
so is that the correlation effects can significantly renormalize
the electronic states of the partially filled bands. Hence the WFs
computed from non-interacting Bloch states are not an optimal
choice for the basis set any more. In Sec.~\ref{WF_DMFT} we
discuss how to calculate within DMFT the local Green function with
the input of the Hamiltonian matrix in WF basis set instead of the
LDA DOS (which is valid only in the case of degenerate bands). In
Sec.~\ref{self-cons} we show that the self-energy operator within
the WF subspace, which is the solution of the correlation problem,
can be transformed back into the full-orbital Hilbert space, thus
enabling the computation of the full interacting Green function
$G(\bf r,\bf r^\prime,\varepsilon)$. It can then be used to
calculate the spectral, magnetic and electronic properties of the
system under investigation. In addition, to make the calculation
scheme fully self-consistent, one can employ the $G(\bf r,\bf
r^\prime,\varepsilon)$ to calculate the correlation affected
charge density and thus the new LDA potential. Thereby the
feedback from DMFT to LDA can be incorporated in a well-defined
way. This is actually one of the great advantages of using the WF
basis since in the LMTO basis the feedback from DMFT to LDA is
essentially uncontrolled. In Sec.~\ref{results} the results for
the electronic structure of the two vanadium oxides SrVO$_3$ and
V$_2$O$_3$ obtained by the method developed in this work are
presented and compared with the previous calculations by the
simpler methods and new bulk-sensitive spectra. Finally in
Sec.~\ref{conclusion} we close this work with a conclusion.

%%%%%%%%%%%%%%%%%%%%%%%%%%%%%%%%%%%%%%%%%%%%%%%%%%%%%%%%%%%%%%%%%%%%%%%%%%%%
% Method
%%%%%%%%%%%%%%%%%%%%%%%%%%%%%%%%%%%%%%%%%%%%%%%%%%%%%%%%%%%%%%%%%%%%%%%%%%%%
\section{Method}
\label{method}

Let us consider the general case of the electronic structure problem. For
the LDA Hamiltonian $\hat H$ we have a Hilbert space of eigenfunctions
(Bloch states $|\psi_{i\bf k}\rangle$) with the basis set $|\phi_\mu\rangle$
defined by particular methods (e.g., LMTO~\cite{LMTO}, or linearized augmented plane
waves (LAPW)~\cite{Mattheiss86}, etc.). In this basis set the Hamiltonian
operator is defined as
%%%%%%%%%%%%%%%%%%%%%%%%%%%%%%%%%%%%%%%%%%%%%%%%%%%%%%%%%%%%%%%%%%%%%%%%%%%%
\begin{eqnarray}
\label{Ham_gen}
 \widehat H & = & \sum_{\mu\nu} |\phi_\mu\rangle
H_{\mu\nu} \langle\phi_\nu|.
\end{eqnarray}
%%%%%%%%%%%%%%%%%%%%%%%%%%%%%%%%%%%%%%%%%%%%%%%%%%%%%%%%%%%%%%%%%%%%%%%%%%%%
Here and later greek indices are used for full-orbital matrices.

If we consider a certain subset of the Hamiltonian eigenfunctions, for example
Bloch states of partially filled bands $|\psi_{n\bf k}\rangle$, we can define
a corresponding subspace in the total Hilbert space. The Hamiltonian matrix is
diagonal in the Bloch states basis. However, physically more appealing is a
basis set which has the form of site-centered atomic orbitals. That is a set
of WFs $|W_{n}^{\bf T}\rangle$ defined as the Fourier transformation of a
certain linear combination of Bloch functions belonging to this subspace (see
below (\ref{WF_psi})). The Hamiltonian operator $\hat H^{WF}$ defined in this
basis set is
%%%%%%%%%%%%%%%%%%%%%%%%%%%%%%%%%%%%%%%%%%%%%%%%%%%%%%%%%%%%%%%%%%%%%%%%%%%%
\begin{eqnarray}
\label{Ham_genWF} \widehat H^{WF} & = & \sum_{nn'{\bf
T}}|W_{n}^{\bf 0}\rangle H_{nn'}({\bf T}) \langle W_{n'}^{\bf T}|.
\end{eqnarray}
%%%%%%%%%%%%%%%%%%%%%%%%%%%%%%%%%%%%%%%%%%%%%%%%%%%%%%%%%%%%%%%%%%%%%%%%%%%%

The total Hilbert space can be divided into a direct sum of the above
introduced subspace (of partially filled Bloch states) and the subspace
formed by all other states orthogonal to it. Those two subspaces are
decoupled since they are the eigenfunctions corresponding
to different eigenvalues. The Hamiltonian matrix in the WF basis (i.e.,
a collection of the bases of the specific subspaces) is block-diagonal
so that the matrix elements between different subspaces
are zero. The block matrix $H_{nn'}$ in (\ref{Ham_genWF}) corresponding
to the partially filled bands can be considered as a projection of the
full-orbital Hamiltonian operator (\ref{Ham_gen}) onto the subspace defined
by its WFs.

All this concerns the non-interacting (or LDA) Hamiltonian. To treat
Coulomb correlations one also needs a definition of the localized orbitals where
the electrons interact. WFs are a natural choice for
such a definition. This choice leads to an important flexibility in the size of
the basis set in the sense that the number of WFs can be changed by changing
the set of Bloch bands considered. The simplest case is a set of
partially filled bands, for example the $t_{2g}$-bands of vanadium oxides.
This is a physically justified approximation because the Coulomb interaction
happens mainly between electrons (or holes) in the partially filled bands.
If the problem to be solved concentrates on the excitation spectrum in a
small energy window around the Fermi level, this basis set is sufficient.
However, if the excitations to higher lying states (real or virtual) are also
important, the set of Bloch bands used to construct the WFs need to be
extended so that the Coulomb interaction will be treated in a larger Hilbert
subspace.

Practically this means that the correlation problem is solved using
a non-interacting few-orbital Hamiltonian $\hat H^{WF}$
(\ref{Ham_genWF}) instead of the full Hilbert space
Hamiltonian $\hat H$ (\ref{Ham_gen}). The interaction matrix elements of the
model Hamiltonian can be determined from constrained LDA
calculations for the specific WF basis set (\ref{U_WF}).

Projecting the full orbital Hilbert space Hamiltonian $\hat H$ (\ref{Ham_gen})
onto the subspace of the partially filled bands gives us a few-orbital Hamiltonian
$\hat H^{WF}$ (\ref{Ham_genWF}). This significantly decreases the complexity
of the correlation problem, thus permitting its explicit solution.
The many-body problem with a Hubbard
interaction then leads to a local self-energy operator
$\widehat\Sigma^{WF}(\varepsilon)$ which is naturaly defined in the basis of WFs
centered on the same site:
%%%%%%%%%%%%%%%%%%%%%%%%%%%
\begin{eqnarray}
\label{Sigma_gen} \widehat{\Sigma}^{WF}(\varepsilon) & = & \sum_{nn'}
|W_{n}^0 \rangle \Sigma_{nn'}(\varepsilon) \langle W_{n'}^0|.
\end{eqnarray}
%%%%%%%%%%%%%%%%%%%%%%%%%%%
We note that, in contrast to other ``basis-reducing'' methods, the
information about the states corresponding to the bands below and
above the projected ones is not lost. In fact, the information is
stored in the ${\bf k}$-dependent projection matrix between the
full orbital basis set and the orthonormalized WFs
(\ref{coef_WF_Bloch}). The definition (\ref{Sigma_gen}) allows one
to convert the self-energy matrix $\Sigma_{nn'}(\varepsilon)$ back
to the full Hilbert space basis set (subsection~\ref{self-cons}).
With this the interacting Green function can also be calculated in
the full-orbital Hilbert space (subsection~\ref{WF_GF}).

%%%%%%%%%%%%%%%%%%%%%%%%%%%%%%%%%%%%%%%%%%%%%%%%%%%%%%%%%%%%%%%%%%%%%%%%%%%%
% Definition and construction of Wannier functions
%%%%%%%%%%%%%%%%%%%%%%%%%%%%%%%%%%%%%%%%%%%%%%%%%%%%%%%%%%%%%%%%%%%%%%%%%%%%
\subsection{Definition and construction of Wannier functions}
\label{W_def}

The concept of WFs has a very important place in the electron theory in solids
since its first introduction in 1937 by Wannier \cite{wannier}.  WFs are the Fourier
transformation of Bloch states $|\psi_{i\bf k}\rangle$
%%%%%%%%%%%%%%%%%%%%%%%%%%%
\begin{eqnarray}
\label{WF_psi_def}
|W_{i}^{\bf T}\rangle& = & \frac{1}{\sqrt{N}}\sum_{\bf k}
e^{-i{\bf kT}}|\psi_{i{\bf k}}\rangle,
\end{eqnarray}
%%%%%%%%%%%%%%%%%%%%%%%%%%%
where $N$ is the number of discrete ${\bf k}$ points in the first Brillouin
zone (or, the number of cells in the crystal). These extremely convenient
orthogonal functions were widely investigated in the seventies
\cite{all_links}. The strongly localized nature of the WFs together with all
advantages of the atomic functions makes them a very useful tool where
the atomic character of the electrons is highlighted. Thus, using the
WF method, significant progress was achieved in the fields of narrow-band
superconductors, disordered systems, solid surfaces, etc. Several methods for
calculating WF for single and multiple bands in periodic crystals, and their
generalization to non-periodic systems were proposed. The problem of
non-unique definition of WFs in these methods was resolved by an iterative
optimization of trial functions which have the same real and point group
symmetry properties as WFs. Among these methods, there are the variational
Koster-Parzen principle \cite{Koster53,Parzen53} which was
generalized by Kohn \cite{K59,K73_2485,K73,K74,K74_448,K93},  the general
pseudopotential formalism proposed by Anderson \cite{Anderson68}, and the
projection operator formalism by Cloizeaux \cite{C63,C64,C64_698}. However,
all these computational schemes are restricted to simple band
structures.

Wannier functions are not uniquely defined because for a certain set of bands
any orthogonal linear combination of Bloch functions $|\psi_{i\bf k}\rangle$
can be used in (\ref{WF_psi_def}). In general it means that the freedom of
choice of Wannier functions corresponds to  freedom of choice of a unitary
transformation matrix $U^{({\bf k})}_{ji}$ for corresponding Bloch functions
\cite{vanderbildt}:
%%%%%%%%%%%%%%%%%%%%%%%%%%%
\begin{eqnarray}
\label{psi_def} |\widetilde\psi_{i\bf k}\rangle & = & \sum_j U^{({\bf
k})}_{ji} |\psi_{j\bf k}\rangle.
\end{eqnarray}
%%%%%%%%%%%%%%%%%%%%%%%%%%%
The resulting Bloch function $|\widetilde\psi_{i\bf k}\rangle$ will generally not be
an eigenfunction of the Hamiltonian but has the meaning of a Bloch
sum of Wannier functions (see below $|\widetilde{W}_{n\bf k}\rangle$ in
(\ref{WF_psi})). There is no rigorous way to define $U^{({\bf k})}_{ji}$.
This calls for an additional restriction on the properties of WFs. Among
others Marzari and Vanderbilt \cite{vanderbildt} proposed the condition of
maximum localization for WFs, resulting in a variational procedure to calculate
$U^{({\bf k})}_{ji}$. To get a good initial guess the authors of \cite{vanderbildt}
proposed choosing a set of localized trial orbitals $|\phi_n\rangle$ and
projecting them onto the Bloch functions $|\psi_{i\bf k}\rangle$. It was
found that this starting guess is usually quite good. This fact later led to
the simplified calculating scheme proposed in \cite{pickett} where the
variational procedure was abandoned and the result of the aforementioned
projection was considered as the final step.
The approach of \cite{vanderbildt} has recently been
used for the investigation of the
row of 3$d$ transition metals (Fe, Co, Ni, and Cu) within the simplest
many-body approximation, namely the unscreened Hartree-Fock approximation
\cite{Schnell03}.

Another possibility to construct WFs was recenly developed by Andersen
$et~al.$ \cite{NMTO}. They proposed the Nth-order muffin-tin
orbital (NMTO) scheme in which Wannier-like low-energy MTOs can be designed
{\it a priori}. Using a new implementation of the LDA+DMFT approach they
performed an investigation of the Mott transition in orthorhombic 3$d^{1}$
perovskites \cite{Pavarini03}. In this approach a realistic Hamiltonian
constructed with Wannier orbitals (on symmetrically orthonormalized NMTOs)
was solved by DMFT, including the non-diagonal part of the on-site
self-energy.

Our projection procedure works as follows.
First of all one needs to identify the physically relevant bands
which will then be projected onto a WF basis.
For example, in perovskites one usually takes the partially filled
$d$-shell or some particular $d$-bands of transition metals, since they are
mainly responsible for the physical properties of the
system~\cite{LDADMFT}.
These orbitals are well-separated and are, in our approach,
easily extracted from the full orbital space as will be shown later.
Moreover, the projection method is applicable even in the case where
the bands of interest differ and are strongly hybridized (for example,
Cu-3$d$ and O-2$p$ states in high-T$_c$ superconductors~\cite{ZETF}).

To project bands of particular symmetry onto the WFs basis one can
select either the band indices of the corresponding Bloch functions
($N_1,...,N_2$), or choose the energy interval ($E_1,E_2$) in which
the bands are located. Non-orthogonalized WFs in reciprocal space
$|\widetilde{W}_{n\bf k}\rangle$ are then the projection of the set of
site-centered atomic-like trial orbitals $|\phi_n\rangle$ on the
Bloch functions $|\psi_{i\bf k}\rangle$ of the chosen bands
(band indices $N_1$ to $N_2$, energy interval ($E_1,E_2$)):
%%%%%%%%%%%%%%%%%%%%%%%%%%%
\begin{eqnarray}
\label{WF_psi} |\widetilde{W}_{n\bf k}\rangle & \equiv & \sum_{i=N_1}^{N_2}
|\psi_{i\bf k}\rangle\langle\psi_{i\bf k}|\phi_n\rangle =
\sum_{i(E_1\le \varepsilon_{i}({\bf k})\le E_2)} |\psi_{i\bf
k}\rangle\langle\psi_{i\bf k}|\phi_n\rangle.
\end{eqnarray}
%%%%%%%%%%%%%%%%%%%%%%%%%%%
Then the real space WFs $|\widetilde{W}_{n}^{\bf T}\rangle$ are given by
%%%%%%%%%%%%%%%%%%%%%%%%%%%
\begin{eqnarray}
\label{WF_real} |\widetilde{W}_{n}^{\bf T}\rangle & = & \frac{1}{\sqrt{N}}
\sum_{\bf k} e^{-i{\bf kT}}|\widetilde{W}_{n\bf k}\rangle.
\end{eqnarray}
%%%%%%%%%%%%%%%%%%%%%%%%%%%
In the present work the trial orbitals $|\phi_n\rangle$ are LMTOs.
Note that in the multi-band case a WF in reciprocal space
$|\widetilde{W}_{n\bf k}\rangle$ does not coincide with the Bloch
function $|\psi_{n\bf k}\rangle$ due to the
summation over band index $i$ in (\ref{WF_psi}). One can consider
them as Bloch sums of WFs analogous to the basis
function Bloch sums $\phi_j^{\bf k}({\bf r})$ (\ref{psik}).

The coefficients $\langle\psi_{i\bf k}|\phi_n\rangle$ in
(\ref{WF_psi}) define (after orthonormalization) the unitary
transformation matrix $U^{({\bf k})}_{ji}$ in (\ref{psi_def}).
However, the projection procedure defined in (\ref{WF_psi}) is
more general than the unitary transformation
(\ref{psi_def}). Namely, the number of bands $(N_2-N_1+1)$ can be larger
than the number of trial functions. In this
case the projection (\ref{WF_psi}) will produce $N$ new functions
$|\widetilde{W}_{n\bf k}\rangle $ which define a certain subspace
of the original $(N_2-N_1+1)$-dimensional space. This subspace
will have the symmetry of the set of trial functions. In the next subsection we
propose a way to determine WFs from the Green
function of the system (\ref{WF1}) rather than from a set of Bloch
states as in (\ref{WF_psi}). In this alternative projection procedure, trial
functions are projected onto the subspace defined by the Green function
in a certain energy interval.

The Bloch functions in LMTO basis (or any other atomic orbital-like
basis set) are defined as
%%%%%%%%%%%%%%%%%%%%%%%%%%%
\begin{eqnarray}
\label{psi} |\psi_{i\bf k}\rangle
& = & \sum_{\mu} c^{\bf k}_{\mu i}|\phi_{\mu}^{\bf k}\rangle,
\end{eqnarray}
%%%%%%%%%%%%%%%%%%%%%%%%%%%
where $\mu$ is the combined index representing $qlm$ ($q$ is the atomic number in
the unit cell, $lm$ are orbital and magnetic quantum numbers),
$\phi_{\mu}^{\bf k}({\bf r})$ are the Bloch sums of the basis orbitals
$\phi_{\mu}({\bf r-T})$
%%%%%%%%%%%%%%%%%%%%%%%%%%%
\begin{eqnarray}
\label{psik} \phi_{\mu}^{\bf k}({\bf r}) & = & \frac{1}{\sqrt{N}}
\sum_{\bf T} e^{i\bf kT} \phi_{\mu}({\bf r}-{\bf T}),
\end{eqnarray}
%%%%%%%%%%%%%%%%%%%%%%%%%%%
and the coefficients have the property
%%%%%%%%%%%%%%%%%%%%%%%%%%%
\begin{equation}
\label{coef} c^{\bf k}_{\mu i}=\langle\phi_{\mu}|\psi_{i \bf k}\rangle.
\end{equation}
%%%%%%%%%%%%%%%%%%%%%%%%%%%

If $n$ in $|\phi_n\rangle$ corresponds to the particular $qlm$
combination (in other words $|\phi_n\rangle$ is an {\it
orthogonal} LMTO basis set orbital), then
$\langle\psi_{i\bf k}|\phi_n\rangle = c_{ni}^{{\bf k}*}$, and hence
%%%%%%%%%%%%%%%%%%%%%%%%%%%
\begin{eqnarray}
\label{WF}
|\widetilde{W}_{n\bf
k}\rangle & = &  \sum_{i=N_1}^{N_2} |\psi_{i\bf k}\rangle
c_{ni}^{{\bf k}*}
 =  \sum_{i=N_1}^{N_2} \sum_{\mu} c^{\bf k}_{\mu i} c_{ni}^{{\bf k}*}
|\phi_{\mu}^{\bf k}\rangle  = \sum_{\mu} \tilde{b}^{\bf k}_{\mu n}
|\phi_{\mu}^{\bf k}\rangle,
\end{eqnarray}
%%%%%%%%%%%%%%%%%%%%%%%%%%%
with
%%%%%%%%%%%%%%%%%%%%%%%%%%%
\begin{equation}
\label{coef_b}
\tilde{b}^{\bf k}_{\mu n}
\equiv  \sum_{i=N_1}^{N_2} c^{\bf k}_{\mu i} c_{ni}^{{\bf k}*}.
\end{equation}
%%%%%%%%%%%%%%%%%%%%%%%%%%%
For a non-orthogonal basis set see Appendix \ref{ortho}.

In order to orthonormalize the WFs (\ref{WF}) one needs to calculate the
overlap matrix $O_{nn'}({\bf k})$
%%%%%%%%%%%%%%%%%%%%%%%%%%%
\begin{eqnarray}
\label{O-S} O_{nn'}({\bf k})&\equiv& \langle\widetilde{W}_{n\bf
k}|\widetilde{W}_{n'\bf k}\rangle = \sum_{i=N_1}^{N_2} c^{\bf k}_{ni}
c_{n'i}^{{\bf k}*},
\end{eqnarray}
%%%%%%%%%%%%%%%%%%%%%%%%%%%
and its inverse square root $S_{nn'}({\bf k})$ is defined as
%%%%%%%%%%%%%%%%%%%%%%%%%%%
\begin{eqnarray}
\label{SS}
S_{nn'}({\bf k}) &\equiv& O^{-1/2}_{nn'}({\bf k}).
\end{eqnarray}
%%%%%%%%%%%%%%%%%%%%%%%%%%%
In the derivation of (\ref{O-S}) the orthogonality of Bloch states
$\langle\psi_{n\bf k}|\psi_{n'\bf k} \rangle=\delta_{nn'}$ was used.

From (\ref{WF}) and (\ref{SS}), the orthonormalized WFs in ${\bf
k}$-space $|W_{n\bf k}\rangle$ can be obtained as
%%%%%%%%%%%%%%%%%%%%%%%%%%%
\begin{eqnarray}
\label{WF_orth} |W_{n\bf k}\rangle = \sum_{n'} S_{nn'}({\bf k})
|\widetilde{W}_{n'\bf k}\rangle
=\sum_{i=N_1}^{N_2} |\psi_{i\bf k}\rangle
\bar{c}_{ni}^{{\bf k}*}  = \sum_{\mu} b^{\bf
k}_{\mu n} |\phi_{\mu} ^{\bf k}\rangle,
\end{eqnarray}
%%%%%%%%%%%%%%%%%%%%%%%%%%%
with
%%%%%%%%%%%%%%%%%%%%%%%%%%%
\begin{eqnarray}
\label{coef_WF_Bloch}
\bar{c}_{ni}^{{\bf k}*}\equiv \langle\psi_{i\bf k}|W_{n\bf k}\rangle
=\sum_{n'} S_{nn'}({\bf k})
c_{n'i}^{{\bf k}*},
\end{eqnarray}
%%%%%%%%%%%%%%%%%%%%%%%%%%%
%%%%%%%%%%%%%%%%%%%%%%%%%%%
\begin{eqnarray}
\label{coef_WF_LMTO}
b^{\bf k}_{\mu n} \equiv \langle\phi_{\mu}^{\bf k}|W_{n\bf k}\rangle=
\sum_{i=N_1}^{N_2} \bar{c}^{\bf k}_{\mu i} \bar{c}_{ni}^{{\bf k}*}.
\end{eqnarray}
%%%%%%%%%%%%%%%%%%%%%%%%%%%

The real space site-centered WFs at the origin $|W_{n}^{\bf
0}\rangle$ are given by the Fourier transform of $|W_{n\bf k}\rangle$ with
${\bf T}=0$. From (\ref{WF_orth}) and (\ref{psik}) one finds
%%%%%%%%%%%%%%%%%%%%%%%%%%%
\begin{eqnarray}
\label{real_WF}  W_n({\bf r})= \frac{1}{\sqrt{N}}\sum_{\bf k}
\langle{\bf r}|W_{n\bf k}\rangle
 & = & \sum_{{\bf T},\mu}\biggl(\frac{1}{N}\sum_{\bf k}
 e^{i\bf kT} b_{\mu n}^{\bf k}\biggr)
\phi_{\mu}({\bf r}-{\bf T}) \nonumber \\
 & = &  \sum_{{\bf T},\mu} w'(n,\mu,{\bf T}) \phi_{\mu}({\bf r}-{\bf T})\\\nonumber
 & = &\sum_{s} w(n,s) \phi_{\alpha(s)}({\bf r}-{\bf T}_s), \\ \nonumber
\end{eqnarray}
%%%%%%%%%%%%%%%%%%%%%%%%%%%
where $w'$ and $w$ are the expansion coefficients of WF in terms of the corresponding
LMTO orbitals, in particular,
%%%%%%%%%%%%%%%%%%%%%%%%%%%
\begin{eqnarray}
w(n,s)  = \frac{1}{N}\sum_{\bf k} e^{i{\bf kT}_s} b_{\alpha(s)n}^{\bf k}.
\end{eqnarray}
%%%%%%%%%%%%%%%%%%%%%%%%%%%
Here $s$ is an index counting the orbitals of the neighboring
cluster for the atom where orbital $n$ is centered (${\bf T}_s$ is
the corresponding translation vector, $\alpha(s)$ is a combined
$qlm$ index). The explicit form of the real space WF
(\ref{real_WF}) can be used to produce, e.g., shapes of chemical bonds.

For other applications only the matrix elements of the various
operators in the basis of WF(\ref{WF_orth}) are needed. From
(\ref{WF_orth}), (\ref{coef_WF_Bloch}) and (\ref{real_WF}) the
matrix elements of the Hamiltonian $\widehat H^{WF}$  in the basis
of WF in real space where both orbitals are in the same unit cell
are
%%%%%%%%%%%%%%%%%%%%%%%%%%%
\begin{eqnarray}
\label{E_WF} H^{WF}_{nn'}({\bf 0}) & = & \langle W_{n}^{\bf
0}|\biggl(\frac{1}{N}\sum_{\bf k}\sum_{i=N_1}^{N_2}|\psi_{i\bf k}\rangle
\epsilon_{i}({\bf k})\langle\psi_{i\bf k}|\biggr)|W_{n'}^{\bf
0}\rangle  \nonumber \\
       & = & \frac{1}{N}
       \sum_{\bf k}\sum_{i=N_1}^{N_2} \bar{c}^{\bf k}_{ni}
      \bar{c}_{n'i}^{{\bf k}*}\epsilon_{i}({\bf k}).
\end{eqnarray}
%%%%%%%%%%%%%%%%%%%%%%%%%%%
$\epsilon_{i}({\bf k})$ is the eigenvalue for a particular band.

If, on the other hand, one of the orbitals corresponds to the WF
for the atom $n'$ shifted from its position in the primary unit
cell by a translation vector ${\bf T}$, then the corresponding
Hamiltonian matrix element is
%%%%%%%%%%%%%%%%%%%%%%%%%%%
\begin{eqnarray}
\label{E_WF2} H^{WF}_{nn'}(\bf T) & = & \langle W_{n}^{\bf 0}|
\hat{H} |W_{n'}^{\bf T}\rangle =
\frac{1}{N} \sum_{\bf k}\sum_{i=N_1}^{N_2}
\bar{c}^{\bf k}_{ni} \bar{c}_{n'i}^{{\bf k}*}
\epsilon_{i}({\bf k}) e^{-i\bf kT}.
\end{eqnarray}
%%%%%%%%%%%%%%%%%%%%%%%%%%%

Matrix elements of the density matrix operator (occupation matrix
$Q^{WF}_{nm}$) in the basis of WFs can be calculated as
%%%%%%%%%%%%%%%%%%%%%%%%%%%
\begin{eqnarray}
\label{Q_WF}
Q^{WF}_{nn'}(\bf T) & = & \langle W_{n}^{\bf
0}|\biggl(\frac{1}{N}\sum_{\bf k}\sum_{i=N_1}^{N_2}|\psi_{i\bf k}\rangle
\theta(E_f - \epsilon_{i}({\bf k}) ) \langle \psi_{i\bf
k}|\biggr)|W_{n'}^{\bf T}\rangle = \nonumber \\
 & = & \frac{1}{N}\sum_{\bf k}\sum_{i=N_1}^{N_2}\bar{c}^{\bf k}_{ni}
\bar{c}_{n'i}^{{\bf k}*}
\theta(E_f-\epsilon_{i}({\bf k}) ) e^{-i\bf kT}
\end{eqnarray}
%%%%%%%%%%%%%%%%%%%%%%%%%%%
$\theta(x)$ is  step function, $E_f$ is the Fermi energy.

Finally, the matrix elements of the  Hamiltonian $\widehat H^{WF}$ in reciprocal
space  are
%%%%%%%%%%%%%%%%%%%%%%%%%%%
\begin{eqnarray}
\label{E_WF_k} H^{WF}_{nn'}({\bf k}) & = & \langle W_{n\bf
k}|\biggl(\frac{1}{N}\sum_{\bf k'}\sum_{i=N_1}^{N_2} |\psi_{i\bf
k'}\rangle \epsilon_{i}({\bf
k'})\langle\psi_{i\bf k'}|\biggr)|W_{n'\bf k}\rangle \nonumber \\
       & = & \sum_{i=N_1}^{N_2} \bar{c}^{\bf k}_{ni} \bar{c}_{n'i}^{{\bf k}*}
       \epsilon_{i}({\bf k}).
\end{eqnarray}
%%%%%%%%%%%%%%%%%%%%%%%%%%%
The (\ref{E_WF_k}) is valid only if the WFs computed by eqs.
(\ref{WF_orth})-(\ref{coef_WF_LMTO}).
If the WFs were obtained in one calculation and then used to compute
the Hamiltonian matrix in another (as is the case for
the WFs (\ref{WF1}) in the Green functions formalism
(see subsection~\ref{WF_GF})) then
eq. (\ref{E_WF_k}) is not valid any more and the general
expression must be used:
\begin{eqnarray}
\label{E_WF3}
 H^{WF}_{nn'}({\bf k}) & = & 
\sum_{i=N_1}^{N_2} \epsilon_{i}({\bf k})  
\sum_{\mu} b_{\mu n}^{{\bf k} *} c_{\mu i}^{\bf k}
\sum_{\nu} b_{\nu n'}^{\bf k} c_{\nu i}^{{\bf k} *}
\end{eqnarray}

Thus, the transformation from LMTO to WF basis set is defined by the explicit
form of WFs (\ref{WF_orth},\ref{coef_WF_LMTO}), and by the expressions for
matrix elements of the Hamiltonian and density matrix operators in WF basis
(\ref{E_WF},\ref{Q_WF}). The back transformation from WF  to LMTO basis can also
be defined using (\ref{WF_orth}) (see subsection~\ref{self-cons}).

Finally, the Coulomb matrix element $U$ needs to be calculated in the same WF basis.
This requires a method similar to constrained LDA~\cite{Gunnarsson89}, but now for WFs.
To this end the WF-energy (\ref{E_WF}) is computed as a function of its
occupancy (\ref{Q_WF}) for a given WF $n$. Then the corresponding Coulomb interaction
parameter $U_n$ in the WF basis is given by
%%%%%%%%%%%%%%%%%%%%%%%%%%%
\begin{eqnarray}
\label{U_WF}
U_n \equiv \frac{d H_{nn}^{WF}({\bf 0})}{d Q_{nn}^{WF}({\bf 0})}.
\end{eqnarray}
%%%%%%%%%%%%%%%%%%%%%%%%%%%
As one can see $U_n$ depends on the WFs via (\ref{E_WF},\ref{Q_WF}).
Once the WFs have been recalculated (for example in some self-consistent loop)
the interaction has to be recalculated as well.

%%%%%%%%%%%%%%%%%%%%%%%%%%%%%%%%%%%%%%%%%%%%%%%%%%%%%%%%%%%%%%%%%%%%%%%%%%%%
% Wannier functions in the Green function formalism
%%%%%%%%%%%%%%%%%%%%%%%%%%%%%%%%%%%%%%%%%%%%%%%%%%%%%%%%%%%%%%%%%%%%%%%%%%%%
\subsection{Wannier functions in the Green function formalism}
\label{WF_GF}

In many-body theory the system is usually not described by Bloch
functions $|\psi_{i\bf k}\rangle$ (\ref{psi}) and their energies
$\epsilon_{i}({\bf k})$ but by the Green function
%%%%%%%%%%%%%%%%%%%%%%%%%%%
\begin{eqnarray}
\label{Green1} G({\bf r},{\bf r'},\varepsilon) & = & \frac{1}{N}
\sum_{\bf k} G^{\bf
k}({\bf r},{\bf r'},\varepsilon)= \frac{1}{N}
\sum_{\bf k} \sum_{\mu \nu}
%\int d{\bf r}\int d{\bf r'}
\phi_{\mu}^{\bf k}({\bf r}) G^{\bf k}_{\mu \nu}(\varepsilon)
\phi^{*\bf k}_{\nu}({\bf r'}).
\end{eqnarray}
%%%%%%%%%%%%%%%%%%%%%%%%%%%
The Green function matrix $G^{\bf k}_{\mu \nu}(\varepsilon)$ is
defined via the non-interacting Hamiltonian matrix $H_{\mu
\nu}({\bf k})$ and the self-energy matrix $\Sigma^{\bf k}_{\mu
\nu}(\varepsilon)$ (\ref{Sigma_full}) as
%%%%%%%%%%%%%%%%%%%%%%%%%%%
\begin{eqnarray}
\label{Green2} G^{\bf k}_{\mu \nu}(\varepsilon) =  (\varepsilon -
\widehat{H}({\bf k}) - \widehat{\Sigma}(\varepsilon,{\bf
k})+i\eta)^{-1}_{\mu \nu}.
\end{eqnarray}
%%%%%%%%%%%%%%%%%%%%%%%%%%%
We define non-orthonormalized WF obtained by projecting  the trial
orbital $\phi_n({\bf r})$ on the Hilbert subspace defined by the
Green function (\ref{Green1}) in the energy interval ($E_1,E_2$),
namely,
%%%%%%%%%%%%%%%%%%%%%%%%%%%
\begin{eqnarray}
\label{WF1} \widetilde{W}_{n{\bf k}}({\bf r}) & = & -\frac{1}{\pi}
Im \int\limits_{E_1}^{E_2} d\varepsilon \int d{\bf r}' G^{\bf k}({\bf
r},{\bf r}',\varepsilon) \phi_n^{\bf k}({\bf r}')
=  \sum_{\mu } \tilde{b}_{\mu n}^{\bf k} \phi_{\mu}^{\bf k}({\bf r}),\\
\nonumber
\end{eqnarray}
%%%%%%%%%%%%%%%%%%%%%%%%%%%
and
%%%%%%%%%%%%%%%%%%%%%%%%%%%
\begin{eqnarray}
\label{coef_green_b}
\tilde{b}_{\mu n}^{\bf k} & \equiv & -\frac{1}{\pi} Im
\int\limits_{E_1}^{E_2} d\varepsilon  G^{\bf k}_{\mu n}(\varepsilon).
\end{eqnarray}
%%%%%%%%%%%%%%%%%%%%%%%%%%%

In the non-interacting case, the self-energy operator
$\hat\Sigma(\varepsilon,{\bf k})$ is absent, and hence we have
%%%%%%%%%%%%%%%%%%%%%%%%%%%
\begin{eqnarray}
\label{Green3} G^{\bf k}_{\mu \nu}(\varepsilon)
  &=&\sum_i \frac{c^{\bf k}_{\mu i} c_{\nu i}^{{\bf k}*}}
{\varepsilon - \epsilon_{i}({\bf k})+i\eta},
\end{eqnarray}
%%%%%%%%%%%%%%%%%%%%%%%%%%%
where $c^{\bf k}_{\mu i}$ are the eigenvectors (\ref{coef}), and
$\epsilon_{i}({\bf k})$  are the eigenvalues of $\widehat{H}({\bf k})$.
Thus $\tilde{b}_{\mu n}^{\bf k}$ in (\ref{coef_green_b}) becomes
%%%%%%%%%%%%%%%%%%%%%%%%%%%
\begin{eqnarray}
\label{b_koef}  \tilde{b}_{\mu n}^{\bf k} =
\sum_{i=N_1}^{N_2}  c^{\bf k}_{\mu i} c_{ni}^{{\bf k}*},
\end{eqnarray}
%%%%%%%%%%%%%%%%%%%%%%%%%%%
where $N_1,N_2$ are the band numbers which correspond to the energy interval
($E_1,E_2$). Since this recovers the result of (\ref{coef_b}),
we demonstrated that our general definition of WFs (\ref{WF1}) via Green
function reduces to that in terms of Bloch functions (\ref{WF}) in
Sec.~\ref{W_def}.

To orthonormalize $\widetilde{W}_{n{\bf k}}({\bf r})$ defined in (\ref{WF1}),
one can just follow the orthonormalizing procedure made in Sec.~\ref{W_def}
(\ref{O-S}- \ref{coef_WF_LMTO}), which will not be repeated here. But it
should be pointed out that in the Green function formalism the overlap
matrix $O_{nn'}({\bf k})$ is defined as
%%%%%%%%%%%%%%%%%%%%%%%%%%%
\begin{eqnarray}
\label{O-S1} O_{nn'}({\bf k}) & = & \langle \widetilde{W}_{n\bf
k}|\widetilde{W}_{n'\bf k}\rangle = \sum_{\mu } \tilde{b}_{\mu
n}^{{\bf k}*} \tilde{b}_{\mu n'}^{\bf k}. \nonumber
\end{eqnarray}
%%%%%%%%%%%%%%%%%%%%%%%%%%%
The occupancy matrix in the orthogonalized WF basis (\ref{WF1}) is defined as
%%%%%%%%%%%%%%%%%%%%%%%%%%%
\begin{eqnarray}
\label{Q_WF4} Q_{nn'}({\bf T})& = & -\frac{1}{\pi} Im
\int\limits_{-\infty}^{E_F} d\varepsilon \int \int d{\bf r} d{\bf r
}' \frac{1}{N}\sum_{\bf k} W^*_{n{\bf k}}({\bf r}) G^{\bf k}({\bf r},{\bf
r}',\varepsilon) W_{n'{\bf k}}({\bf r}')e^{-i\bf kT}.
\end{eqnarray}
%%%%%%%%%%%%%%%%%%%%%%%%%%%
By using (\ref{Green1}) and orthogonalized (\ref{WF1}), one finds
%%%%%%%%%%%%%%%%%%%%%%%%%%%
\begin{eqnarray}
Q_{nn'}({\bf T})=
\frac{1}{N}\sum_{\bf k} \sum_{\mu \nu} b_{\mu n}^{{\bf k}*}  b_{\nu n'}^{\bf k}
  Q^{\bf k}_{\mu\nu}e^{-i\bf kT},
\end{eqnarray}
%%%%%%%%%%%%%%%%%%%%%%%%%%%
with
%%%%%%%%%%%%%%%%%%%%%%%%%%%
\begin{eqnarray}
 Q^{\bf k}_{\mu\nu} & = & -\frac{1}{\pi} Im \int\limits_{-\infty}^{E_F}
d\varepsilon G^{\bf k}_{\mu \nu}(\varepsilon).
\end{eqnarray}
%%%%%%%%%%%%%%%%%%%%%%%%%%%

The energy matrix  can be defined similarly (except that the
integral over energy is calculated in the $(E_1,E_2)$ interval
where the corresponding WFs are defined) as:
%%%%%%%%%%%%%%%%%%%%%%%%%%%
\begin{eqnarray}
\label{E_WF4} E_{nn'}(\bf T)& = & -\frac{1}{\pi} Im
\int\limits_{E_1}^{E_2}\varepsilon d\varepsilon \int \int d{\bf r}
d{\bf r}' \frac{1}{N}\sum_{\bf k} W^*_{n{\bf k}}({\bf r}) G^{\bf
k}({\bf
r},{\bf r}',\varepsilon) W_{n'{\bf k}}({\bf r}')e^{-i\bf kT}  \\
&=&
\frac{1}{N}
\sum_{\bf k} \sum_{\mu \nu} b_{\mu n}^{{\bf k}*}  b_{\nu n'}^{\bf k}
  E^{\bf k}_{\mu\nu}e^{-i\bf kT},  \nonumber
\end{eqnarray}
with
\begin{eqnarray}
 E^{\bf k}_{\mu\nu} & = & -\frac{1}{\pi} Im \int\limits_{E_1}^{E_2}
\varepsilon d\varepsilon G^{\bf k}_{\mu \nu}(\varepsilon).
\end{eqnarray}
%%%%%%%%%%%%%%%%%%%%%%%%%%%
While (\ref{E_WF4}) looks similar to the non-interacting
Hamiltonian in WFs basis (\ref{E_WF2}), it includes correlations
via $\widehat{\Sigma}(\varepsilon)$ in (\ref{Green2}) and hence
is {\it {interacting}}.

%%%%%%%%%%%%%%%%%%%%%%%%%%%%%%%%%%%%%%%%%%%%%%%%%%%%%%%%%%%%%%%%%%%%%%%%%%%%
% DMFT in the Wannier function formalism
%%%%%%%%%%%%%%%%%%%%%%%%%%%%%%%%%%%%%%%%%%%%%%%%%%%%%%%%%%%%%%%%%%%%%%%%%%%%
\subsection{DMFT in the Wannier function formalism}
\label{WF_DMFT}

In the previous subsection we showed (eqs. \ref{Green1}-\ref{WF1})
that the self-energy operator
is needed to construct the WFs in terms of the full interacting Green function.
The DMFT~\cite{vollha93,pruschke,georges96} was recently found to
be a powerful tool to numerically solve multi-band Hubbard models.
To define parameters of the correlated model Hamiltonian
(hoppings, screened Coulomb integrals), density functional theory
within the LDA was used~\cite{LDADMFT1}.
The combined LDA+DMFT computational scheme was successfully
applied to a wide range of compounds with degenerate
(or almost degenerate) orbitals (for more details see
~\cite{LDADMFT}). In these cases
the non-interacting LDA DOS was used to obtain the Green
function of the system through a Hilbert transformation.
Furthermore, the screened Coulomb interaction parameters U and J
were calculated by constrained LDA~\cite{Gunnarsson89}.

Quite generaly, this scheme needs to be improved in two respects:
(i) instead of the LDA DOS an LDA-Hamiltonian with a few, relevant
orbitals should be
used to calculate the Green function, and (ii) a feedback from DMFT to LDA
should be incorporated.
Both of these problems are solved by the new approach proposed in this work.
In this method the Hamiltonian matrix in the WF basis set
$H^{WF}_{nn'}({\bf k})$ is calculated from the LDA Hamiltonian via
the projection procedure (\ref{WF_psi},\ref{E_WF_k}). In the
DMFT self-consistency loop the local Green function
$G^{loc}_{nn'}(\varepsilon)$ is then computed as an integral over the first
Brillouin zone (BZ) :
%%%%%%%%%%%%%%%%%%%%%%%%%%%
\begin{eqnarray}
\label{gloc} G^{loc}_{nn'}(\varepsilon)=\!\frac{1}{V_{BZ}} \int
d{\bf k} \left( \left[ \;(\varepsilon+E^{(N)}_f)
\widehat{1}-\widehat{H}^{WF}({\bf k})
-\widehat{\Sigma}^{WF}(\varepsilon)\right]^{-1}\right)_{nn'}.
\end{eqnarray}
%%%%%%%%%%%%%%%%%%%%%%%%%%%
The integration can actually be restricted to the irreducible part
of the BZ via the analytical tetrahedron method~\cite{ATM} with a
subsequent symmetrization of the Green function matrix. The
chemical potential $E^{(N)}_f$ is determined by the number of
electrons on the $N$ interacting orbitals of interest
\cite{efermi}.

The DMFT is based on the fact that in the $d=\infty$ limit the self-energy
operator is local~\cite{MetzVoll89,MueHartmann89}. Its matrix
$\Sigma_{nn'}(\varepsilon)$ ($n,n'$ - WF indices) is defined in WF basis
(\ref{Sigma_gen}). If the trial functions in (\ref{WF_psi},\ref{WF1}) are
chosen as the basis functions of the irreducible representation of the point
symmetry group of some particular real system~\cite{O_h_group},
the Green function matrix (\ref{Green2}) and hence the
self-energy matrix (\ref{Sigma_gen}) can be made
diagonal~\cite{nearly} in the $n$ index for
on-site matrix elements.

To set up the DMFT equations one needs to define the bath Green function
${\cal G}$ in the usual way via the Dyson equation\cite{georges96}:
%%%%%%%%%%%%%%%%%%%%%%%%%%%
\begin{eqnarray}
\label{G0} {\cal G}^{-1} = (G^{loc})^{-1} + \Sigma.
\end{eqnarray}
%%%%%%%%%%%%%%%%%%%%%%%%%%%
To obtain the Green function $G^{imp}$ of the effective single
impurity Anderson problem, various methods can be used: quantum
Monte-Carlo (QMC), numerical renormalization group (NRG), exact
diagonalization (ED), non-crossing approximation (NCA), etc. (for
a brief overview of the methods see \cite{LDADMFT}). With the
condition $G^{loc}=G^{imp}$ one closes the self-consistent loop
which can then be iterated until a converged solution for the self-energy
$\Sigma_{nn}(\varepsilon)$ is found.

%%%%%%%%%%%%%%%%%%%%%%%%%%%%%%%%%%%%%%%%%%%%%%%%%%%%%%%%%%%%%%%%%%%%%%%%%%%%
% Converting back to the full-orbital Hilbert space
%%%%%%%%%%%%%%%%%%%%%%%%%%%%%%%%%%%%%%%%%%%%%%%%%%%%%%%%%%%%%%%%%%%%%%%%%%%%
\subsection{Converting back to the full-orbital Hilbert space}
\label{self-cons}

The self-energy operator $\widehat\Sigma^{WF}(\varepsilon)$
obtained as a solution of DMFT in Sec~\ref{WF_DMFT} is
defined in the WF basis set (\ref{Sigma_gen}).
In order to compute the interacting Green function in
the full-orbital Hilbert space (\ref{Green1}-\ref{Green2}) one has
to convert it back to the full-orbital (LMTO) basis set.
This can be easily done by using the
linear expansion form of the WFs in terms of the full-orbital basis set
(\ref{WF_orth},\ref{coef_WF_LMTO}),
%%%%%%%%%%%%%%%%%%%%%%%%%%%
\begin{eqnarray}
\label{Sigma_full}
\Sigma^{\bf
k}_{\mu\nu}(\varepsilon)  =  \langle \phi_{\mu}^{\bf k}|
\widehat{\Sigma}(\varepsilon)|\phi_{\nu}^{\bf k}\rangle
 =  \sum_n  \langle \phi_{\mu}^k|W_{n\bf k} \rangle
\Sigma_{nn'}(\varepsilon)    \langle W_{n'\bf k}|\phi_{\nu}^{\bf
k}\rangle = \sum_n b_{\mu n}^{\bf k} \Sigma_{nn'}(\varepsilon)
b_{\nu n'}^{{\bf k}*}.
\end{eqnarray}
%%%%%%%%%%%%%%%%%%%%%%%%%%%
Here we use the local form of the self-energy operator as obtained in
DMFT, but the formalism can be easily generalized.
In the following we refer to this self-energy operator as the ``full-orbital''
self-energy.

The matrix elements of the self-energy operator $\Sigma^{\bf k}_{\mu\nu}
(\varepsilon)$ (\ref{Sigma_full}) together with the non-interacting
Hamiltonian matrix $H^{\bf k}_{\mu\nu}$ allow one to calculate the
Green function matrix $G^{\bf k}_{\mu\nu}(\varepsilon)$ (\ref{Green2})
and thus the full-orbital interacting Green function
$G({\bf r},{\bf r'},\varepsilon)$ (\ref{Green1}).
$G({\bf r},{\bf r'},\varepsilon)$ contains the full information about
the system, and various electronic, magnetic and spectral properties can
be obtained from it. In Sec.~\ref{full-orbital} we use the full-orbital
interacting Green function computed within DMFT(QMC) to calculate the
photoemission and X-ray absorption spectra for the strongly correlated
vanadium oxides SrVO$_3$ and V$_2$O$_3$, and to compare them with new
bulk-sensitive experimental spectra.

One can also calculate the charge density distribution modified by
correlation effects via
%%%%%%%%%%%%%%%%%%%%%%%%%%%
\begin{eqnarray}
\label{rho}
 \rho(\bf r)&=& -\frac{1}{\pi}Im
\int\limits_{-\infty}^{E_F}d\varepsilon G({\bf r},{\bf
r},\varepsilon).
\end{eqnarray}
%%%%%%%%%%%%%%%%%%%%%%%%%%%
With this $\rho(\bf r)$ one can recalculate the LDA-potential
(which is a functional of electron density). From the full-orbital
Green function (\ref{Green1}) one can recalculate new WFs
(\ref{WF1},\ref{coef_green_b}) which together with the new LDA
Hamiltonian allows one to obtain new parameters for the
non-interacting Hamiltonian (\ref{E_WF3}). With (\ref{U_WF}) one
can then compute a new Coulomb interaction parameter $U$. The set
of new LDA potential, WFs and Coulomb interaction parameters
calculated from the interacting Green function (\ref{Green1})
defines the input for the next iteration step and hence closes the
self-consistency loop in the proposed computation scheme. For the
feedback from DMFT to LDA in the particular case of the LMTO
method \cite{LMTO} one needs a set of moments for the partial
densities of states $M^{(m)}_{ql}$ for every atomic sphere $q$ and
the orbital moment $l$~\cite{Skriver84} in order to calculate the
new charge density and hence the new LDA potential:
%%%%%%%%%%%%%%%%%%%%%%%%%%%
\begin{eqnarray}
\label{M-G}
 M^{(m)}_{ql} & = & \int\limits_{-\infty}^{E_F}d\varepsilon
\varepsilon^m N_{ql}(\varepsilon), \\ \nonumber
N_{ql}(\varepsilon)&=& -\frac{1}{\pi N}Im \sum_k \sum_m
G^k_{qlm,qlm}(\varepsilon).
\end{eqnarray}
%%%%%%%%%%%%%%%%%%%%%%%%%%%

%%%%%%%%%%%%%%%%%%%%%%%%%%%%%%%%%%%%%%%%%%%%%%%%%%%%%%%%%%%%%%%%%%%%%%%%%%%%
% Summary of the WF scheme
%%%%%%%%%%%%%%%%%%%%%%%%%%%%%%%%%%%%%%%%%%%%%%%%%%%%%%%%%%%%%%%%%%%%%%%%%%%%
\subsection{Summary of the WF scheme}
\label{sum_method}

%%%%%%%%%%%%%%%%%%%%%%%%%%%%%%%%%%%%%%%%%%%%%%%%%%%%%%%%%%%%%%%%%%%%%%%%%%%%%
\begin{figure}
\centering
\includegraphics[clip=true,width=1.0\textwidth]{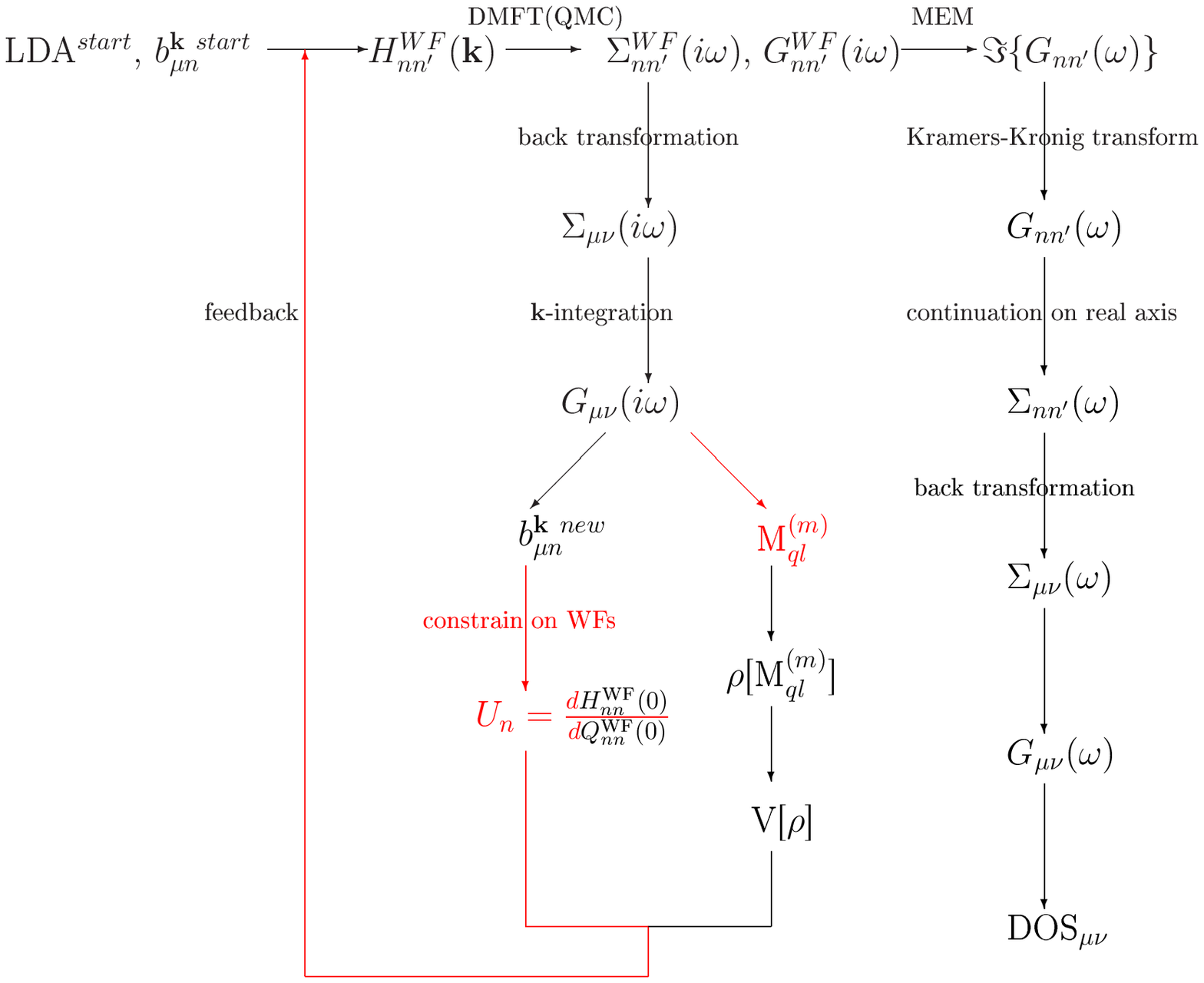}
\caption{Scheme of {\it{ab initio}} fully self-consistent LDA+DMFT
scheme based on the WF formalism (see text). Red color marks steps
that were not performed in this paper.}
\label{scheme}
\end{figure}
%%%%%%%%%%%%%%%%%%%%%%%%%%%%%%%%%%%%%%%%%%%%%%%%%%%%%%%%%%%%%%%%%%%%%%%%%%%%%

For clarity, in Fig.~\ref{scheme} the essential steps of the
WF scheme presented here are summarized. There are four
interconnected parts in this scheme: (i) the basis of WF, (ii) the
matrix elements of the Hamiltonian and the self-energy in WF
basis, (iii) the Coulomb interaction between electrons on the WFs,
and (iv) the projection into the few-orbital basis and back
transformation to the full-orbital basis which retains the
information about all orbitals. First the matrix elements of the
non-interacting Hamiltonian in reciprocal space $H^{WF}_{nn'}({\bf
k})$ (\ref{E_WF_k}) and the interaction term
$\Sigma_{nn'}({i\omega})$ (\ref{Sigma_gen}) are written in the
basis of explicitly defined WFs $|W_{n\bf k}\rangle$
(\ref{WF_orth}). The actual correlation problem, defined by the
sum of these two terms (\ref{gloc}), is then solved within the
LDA+DMFT(QMC) approach~\cite{LDADMFT}. The local self-energy
$\Sigma_{nn'}({i\omega})$ obtained thereby is then transformed
back from the Wannier basis to the full-orbital space (see
subsection~\ref{self-cons}). Furthermore, with the full-orbital
self-energy $\Sigma_{\mu\nu}({i\omega})$ (\ref{Sigma_full}) the
full-orbital Green function $G_{\mu\nu}({i\omega})$
(\ref{Green1},\ref{Green2}) for the correlated electrons is
calculated by a ${\bf k}$-integration over the Brillouin zone
\cite{ATM}. The Green function allows one to determine a new charge
density distribution (\ref{rho}) (in the LMTO case see
(\ref{M-G})) and a new set of WFs (via
(\ref{WF1},\ref{coef_green_b})). This is used to construct a new
LDA-potential and new non-interacting Hamiltonian. Together with
the new Coulomb interaction parameters $U_n$ (\ref{U_WF}) they serve
as the input for the next iteration of calculations, thus
completing the self-consistency loop. It should be stressed that
in this scheme self-consistency involves not only the self-energy
but also the basis of WFs in which it is defined, the charge density
and LDA potential used for constructing the non-interacting Hamiltonian,
and the interaction strength between electrons in the WFs.

After convergency is reached the maximum entropy method
(MEM)~\cite{MEM} can be used to obtain spectral functions. Then,
using the Kramers-Kronig transform, the Green function on real
axis $G_{nn'}({\omega})$ is computed. With that one can construct
the WFs basis self-energy $\Sigma_{nn'}({\omega})$ (see Appendix
\ref{acSE}). By the back transformation of
$\Sigma_{nn'}({\omega})$ with a subsequent ${\bf k}$-integration
the full-orbital Green function $G_{\mu\nu}({\omega})$ is
obtained, which now contains information not only about the states
for which correlations are considered, but also for all other
orbitals of the system. $G_{\mu\nu}({\omega})$ is used to obtain
orbitally resolved spectral functions. This allows one, for
example, to investigate the influence of the correlated orbitals
(e.g., partially filled $t_{2g}$ orbitals of V in SrVO$_3$ and
V$_2$O$_3$) on other orbitals (oxygen $2p$ and Vanadium $e_g$
orbitals). It also makes possible the computation of spectral
functions in a wide energy region and not only in the vicinity of
the Fermi level.

%%%%%%%%%%%%%%%%%%%%%%%%%%%%%%%%%%%%%%%%%%%%%%%%%%%%%%%%%%%%%%%%%%%%%%%%%%%%
% Results and discussion
%%%%%%%%%%%%%%%%%%%%%%%%%%%%%%%%%%%%%%%%%%%%%%%%%%%%%%%%%%%%%%%%%%%%%%%%%%%%
\section{Results and discussion}
\label{results}

In our earlier LDA+DMFT calculation scheme the specific
properties of a material entered only via the LDA partial
densities of states (DOSs) for the orbitals
of interest. This procedure
is valid for systems where the bands of interest are degenerate
(as in cubic crystals; see Appendix~\ref{hilbert}). For more complicated systems
with lower symmetry one needs to employ the scheme proposed in this work,  where
the non-interacting Hamiltonian $\hat H^{WF}({\bf k})$~(\ref{E_WF_k}) (projected
on WFs) describing the $N$ orbitals under consideration is used for the
calculation of $G_{nn'}^{loc}(\varepsilon)$~(\ref{gloc}) within DMFT (see
section~\ref{WF_DMFT}).

In this section we present results of LDA+DMFT calculations using
the projected Hamiltonian $\hat H^{WF}({\bf k})$. The scheme was
applied to SrVO$_3$ which has a cubic perovskite
crystal structure, and to the trigonally distorted
V$_2$O$_3$ (both in the insulating and metallic phase). The results are
compared with previous LDA+DMFT calculations where the Hilbert
transformation of the LDA DOSs was used for
SrVO$_3$~\cite{srvo} (see section~\ref{comp}).
One should note that the DOSs used for DMFT
calculations of V$_2$O$_3$ were obtained by the TB-LMTO-ASA code v.
47~\cite{LMTO} in contrast to~\cite{V2O3PRL} where the DOSs were
calculated by the ASW method~\cite{ASW}. However, the DOSs obtained
in both methods are very similar and do not produce much different
LDA+DMFT results.

The full-orbital calculation scheme proposed in this work allows
one to answer an important question: how do Coulomb correlations
between some orbitals affect the other orbitals, and in
particular, how does the interaction between the partially filled
$t_{2g}$ orbitals of the V3$d$-shell influence the filled O-2$p$
and the unoccupied V-3$d$($e_g^\sigma$) bands? To answer these
questions the matrix of the self-energy operator
$\Sigma_{nn'}(\varepsilon)$ (\ref{Sigma_gen}) was converted back
to the full-orbital basis set from the WF basis, and the full
orbital interacting Green function (\ref{Green2}) was calculated.
>From that, total and partial DOSs were computed to produce
theoretical spectra for comparison with the experimental
photoemission and X-ray absorption data.

In this work, QMC simulations~\cite{QMC} were used to
solve the effective single impurity Anderson problem in the DMFT loop. The
result of the DMFT(QMC) calculation is the self-energy on imaginary energy
axis $\Sigma_{nn'}(i\omega)$. To find the full-orbital self-energy
$\Sigma^{\bf k}_{\mu\nu}(\varepsilon)$ on the real energy axis one
has to perform an analytical continuation. The procedure is described in
Appendix~\ref{continuation}. A self-consistent computation of the charge density as
described in  section~\ref{self-cons} was not yet performed.
Investigations of correlation effects on the unoccupied cubic e$^\sigma_g$ states
for both SrVO$_3$ and V$_2$O$_3$ are also a matter of future calculations.

%%%%%%%%%%%%%%%%%%%%%%%%%%%%%%%%%%%%%%%%%%%%%%%%%%%%%%%%%%%%%%%%%%%%%%%%%%%%
% Comparison of DMFT results
%%%%%%%%%%%%%%%%%%%%%%%%%%%%%%%%%%%%%%%%%%%%%%%%%%%%%%%%%%%%%%%%%%%%%%%%%%%%
\subsection{Comparison of DMFT results obtained by Hilbert
transformation of the LDA DOS and the projected Hamiltonian $\hat H^{WF}
({\bf k})$}
\label{comp}

%%%%%%%%%%%%%%%%%%%%%%%%%%%%%%%%%%%%%%%%%%%%%%%%%%%%%%%%%%%%%%%%%%%%%%%%%%%%%
\begin{figure}[!h]
\centering
\includegraphics[clip=true,width=0.5\textwidth,angle=270]{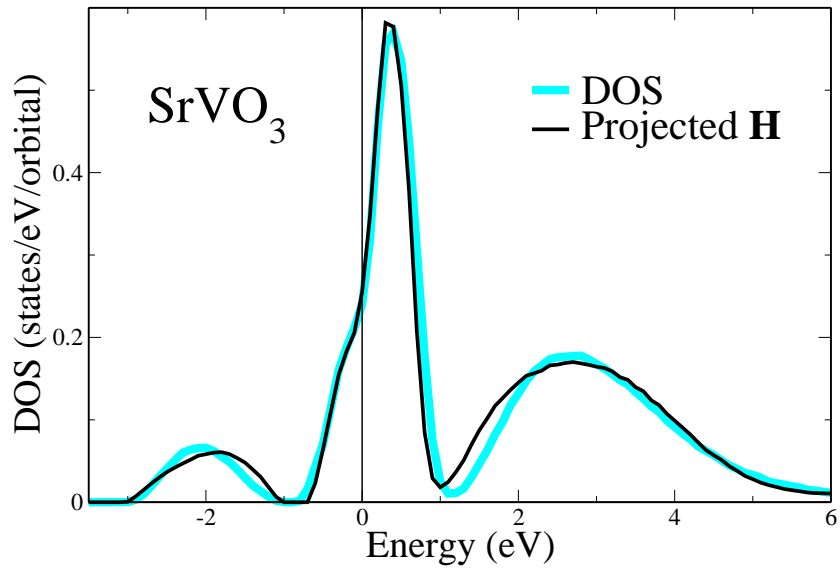}
\caption{Comparison of V-3$d$(t$_{2g}$) spectral functions for SrVO$_3$ calculated
via LDA+DMFT(QMC) using: LDA DOS - light line, projected Hamiltonian - black
line. Fermi level corresponds to zero.}
\label{DOS_HMLT_SrVO3}
\end{figure}
%%%%%%%%%%%%%%%%%%%%%%%%%%%%%%%%%%%%%%%%%%%%%%%%%%%%%%%%%%%%%%%%%%%%%%%%%%%%%

The cubic perovskite SrVO$_3$ can serve as a test case for our new
method. For the cubic $O_h$ point symmetry group, the three 3$d$-orbitals
$xy,xz,yz$ transform according to the triply degenerate $t_{2g}$
irreducible representation of this group. Hence the corresponding
Green function and self-energy matrices have diagonal form with
equal diagonal elements. As shown in Appendix
\ref{hilbert}, the results of the Hilbert transformation using LDA DOS
must coincide with the results of the procedure of integration
over the Brillouin zone with the projected 3x3 t$_{2g}$ Hamiltonian.

We use the same LDA DOS and interaction parameters as in our
previous papers~\cite{srvo}. The V-3$d$(t$_{2g}$) states form a
partially filled band in SrVO$_3$. All t$_{2g}$ orbitals ($xy$,
$xz$, $zy$) are equivalent due to the cubic symmetry of the lattice,
so only the results for one of the t$_{2g}$ orbitals are presented in
Fig.~\ref{DOS_HMLT_SrVO3}. In this figure
V-3$d$(t$_{2g}$) spectral functions, calculated using the LDA DOS and
the projected Hamiltonian are shown. It is easy to see that both results are
almost identical. The small differences between these two
curves are due to the MEM~\cite{MEM}
used for the calculation of the spectral function on the
real energy axis from the DMFT Green function.

We also applied our Hamiltonain scheme to a more complicated system
with lower symmetry where one can expect deviations
from the results obtained using the LDA DOS. We performed
LDA+DMFT(QMC) calculations for the insulating and metallic phases of
V$_2$O$_3$ with the projected 3x3 t$_{2g}$ Hamiltonian and
several $U$ values (the values of $U$ are the same as in~\cite{V2O3PRL}).
In contrast to SrVO$_3$, V$_2$O$_3$ has a non-cubic trigonal symmetry
(space group R$\bar 3$c). Therefore we use the basis functions of the trigonal
$D_{3d}$ point group instead of the cubic $O_h$ group. In this basis set
the V-3$d$-shall is split into two groups of bands. The three t$_{2g}$ bands
are located around the Fermi energy, the two degenerate e$_{g}^\sigma$
bands are at higher energies. Interesting to us are the partially filled
t$_{2g}$ bands which are formed by one $a_{1g}$ and two degenerate
$e_g^{\pi}$ orbitals.

\begin{figure}[!h]
\centering
\includegraphics[clip=true,width=0.7\textwidth]{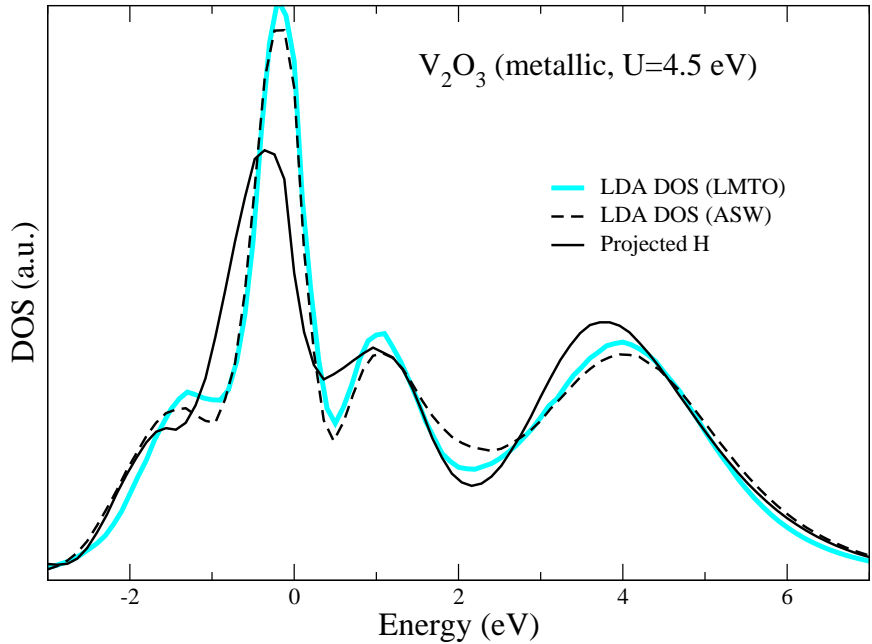}
\caption{Comparison of V-3$d$(t$_{2g}$) spectral functions for V$_2$O$_3$ in
the metallic phase calculated via LDA+DMFT(QMC) using: LDA DOS (LMTO) -
light line~\protect\cite{V2O3PRL}, LDA DOS (ASW) - dashed line,
projected Hamiltonian - black line. Fermi level corresponds to zero.}
\label{DOS_ASW_HMLT_V2O3}
\end{figure}

In Fig.~\ref{DOS_ASW_HMLT_V2O3}, we present V-3$d$(t$_{2g}$) spectral
functions resulting from DOS (LMTO and ASW) and Hamiltonian calculations
at U$=4.5$eV, averaged over the three t$_{2g}$ bands. The spectrum
computed with ASW input is taken from Ref.~\onlinecite{V2O3PRL}.

%%%%%%%%%%%%%%%%%%%%%%%%%%%%%%%%%%%%%%%%%%%%%%%%%%%%%%%%%%%%%%%%%%%%%%%%%%%%%
%%%%%%%%%%%%%%%%%%%%%%%%%%%%%%%%%%%%%%%%%%%%%%%%%%%%%%%%%%%%%%%%%%%%%%%%%%%%%

As mentioned before, the differences between the curves calculated with ASW and
LMTO DOS as input are small; there are only minor deviations in the peak height.
In the comparison between the DOS calculations and the Hamiltonian calculation,
it is interesting to note that for these averaged spectra, the differences are
relatively small. We find the typical four peak structure (with lower Hubbard,
quasiparticle peak and double-peaked upper Hubbard band split by Hund's rule
coupling) for all three calculations. Only the position
and the height of the quasiparticle peak are quite different for the Hamiltonian
calculation, indicating a more insulating solution than for the DOS calculations
at the same U-value. The similarity between the results for the DOS and
Hamiltonian calculations is not surprising, because the trigonal
distortion in V$_2$O$_3$ is relatively small. Therefore the
center of gravity and the bandwidth of the three t$_{2g}$ bands do
not change much and the DOS calculations can still produce accurate results.

\begin{figure}[!h]
\centering
\includegraphics[clip=true,width=0.6\textwidth,angle=270]{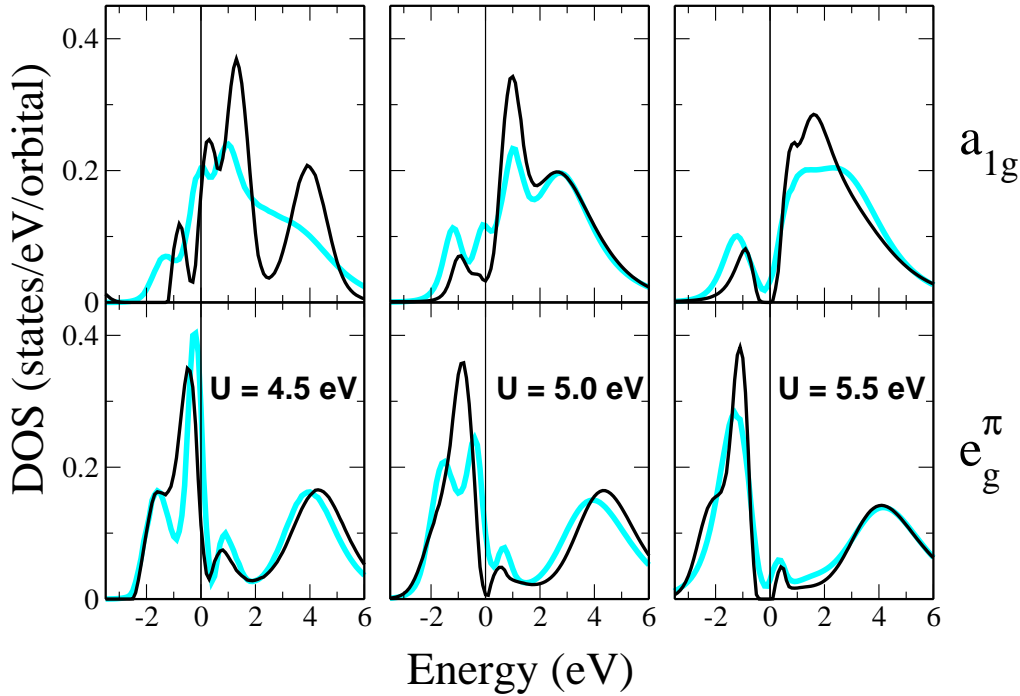}
\caption{Comparison of V-3$d$(t$_{2g}$) spectral functions for V$_2$O$_3$ in
the insulating phase calculated via LDA+DMFT(QMC) using: LDA DOS - light line,
projected Hamiltonian - black line. Upper figures - $a_{1g}$, lower figures
- $e_g^{\pi}$ orbitals. Fermi level corresponds to zero.}
\label{DOS_HMLT_V2O3i}
\end{figure}

\begin{figure}[!h]
\centering
\includegraphics[clip=true,width=0.6\textwidth,angle=270]{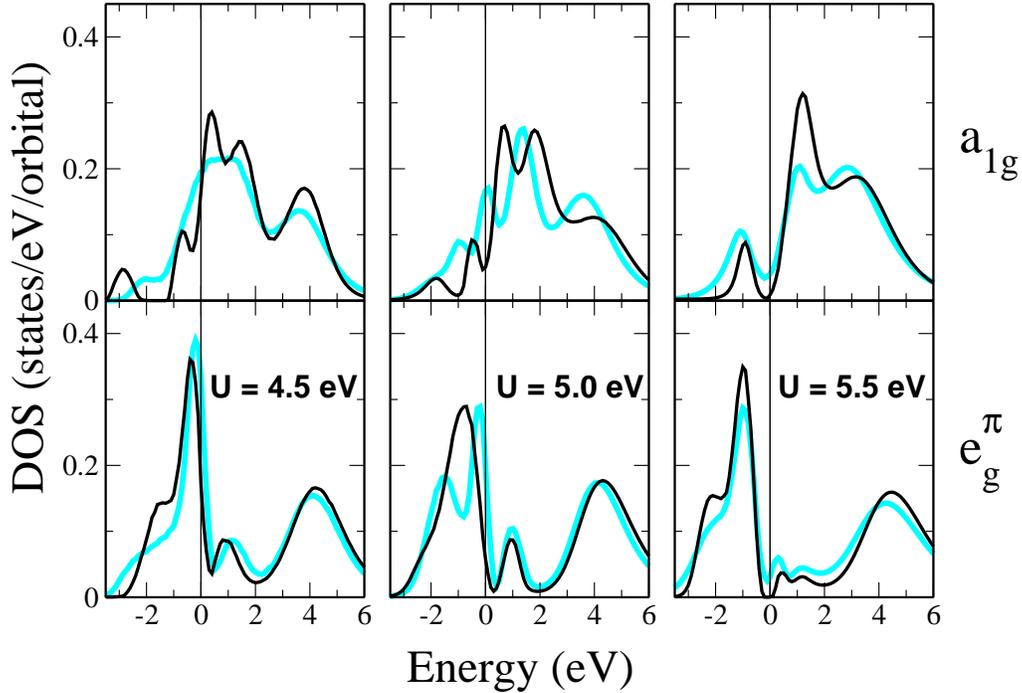}
\caption{Comparison of V-3$d$(t$_{2g}$) spectral functions for V$_2$O$_3$ in
the metal phase calculated via LDA+DMFT(QMC) using: LDA DOS - light line,
projected Hamiltonian - black line. Upper figures - $a_{1g}$, lower figures
- $e_g^{\pi}$ orbitals. Fermi level corresponds to zero.}
\label{DOS_HMLT_V2O3m}
\end{figure}

The differences between DOS and Hamilton calculations are more pronounced when
one compares the band-resolved t$_{2g}$ spectra. In Fig.~\ref{DOS_HMLT_V2O3i}
and Fig.~\ref{DOS_HMLT_V2O3m}, the a$_{1g}$ and e$_g^\pi$ spectra for the
insulating and metallic crystal structure are presented for U$=4.5, 5.0$
and $5.5$ eV.

%%%%%%%%%%%%%%%%%%%%%%%%%%%%%%%%%%%%%%%%%%%%%%%%%%%%%%%%%%%%%%%%%%%%%%%%%%%%%
%%%%%%%%%%%%%%%%%%%%%%%%%%%%%%%%%%%%%%%%%%%%%%%%%%%%%%%%%%%%%%%%%%%%%%%%%%%%%

%%%%%%%%%%%%%%%%%%%%%%%%%%%%%%%%%%%%%%%%%%%%%%%%%%%%%%%%%%%%%%%%%%%%%%%%%%%%%
%%%%%%%%%%%%%%%%%%%%%%%%%%%%%%%%%%%%%%%%%%%%%%%%%%%%%%%%%%%%%%%%%%%%%%%%%%%%%

Here the curves calculated by different schemes are distinctively
different. This is especially clear from the upper part of the
figures~\ref{DOS_HMLT_V2O3i}, \ref{DOS_HMLT_V2O3m} corresponding
to the $a_{1g}$-orbital of the V-3$d$(t$_{2g}$) subband. This is
due to the fact that the hybridization effects are better
accounted for by the projected Hamiltonian, which includes not
only intra-band hoppings but also inter-band on-site and
inter-site hoppings. The latter effectively increases the
bandwidth of the $a_{1g}$-orbital. The corresponding spectral
functions have a different peak structure and different
intensities.

Another conclusion from this series of figures is that the results obtained
with the Hamiltonian procedure show a more insulating behavior for the same
values of $U$ than the results with the scheme using the LDA DOS.

%%%%%%%%%%%%%%%%%%%%%%%%%%%%%%%%%%%%%%%%%%%%%%%%%%%%%%%%%%%%%%%%%%%%%%%%%%%%
% Results of full-orbital calculations
%%%%%%%%%%%%%%%%%%%%%%%%%%%%%%%%%%%%%%%%%%%%%%%%%%%%%%%%%%%%%%%%%%%%%%%%%%%%
\subsection{Results of full-orbital calculations: comparison of calculated
spectra with experimental PES and XAS data}
\label{full-orbital}

%%%%%%%%%%%%%%%%%%%%%%%%%%%%%%%%%%%%%%%%%%%%%%%%%%%%%%%%%%%%%%%%%%%%%%%%%%%%%
\begin{figure}
\centering
\includegraphics[clip=true,width=0.5\textwidth,angle=270]{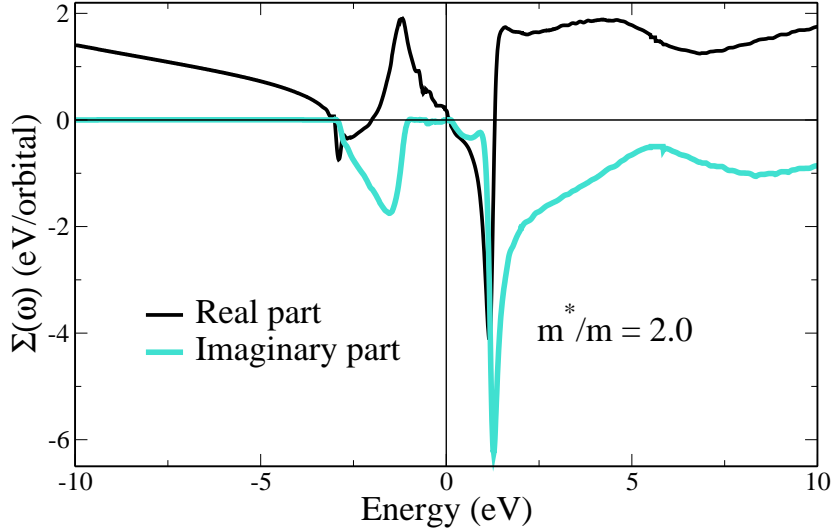}
\caption{Self-energy on real energy axis for SrVO$_3$. Real part - full black
line, imaginary part - full light line. Fermi level corresponds to zero.}
\label{SrVO3_sigma}
\end{figure}
%%%%%%%%%%%%%%%%%%%%%%%%%%%%%%%%%%%%%%%%%%%%%%%%%%%%%%%%%%%%%%%%%%%%%%%%%%%%%

The full-orbital calculations scheme proposed in this work produces an
interacting Green function $G({\bf r},{\bf r'},\varepsilon)$ (\ref{Green1}).
The knowledge of the full Green function allows us to calculate the spectral
function not only for the states with correlations (V-3$d$($t_{2g}$)
orbitals in the case of SrVO$_3$ and V$_2$O$_3$), but also their effect on
the lower lying occupied oxygen 2$p$ states and the higher lying unoccupied
V-3$d$($e_{g}^\sigma$) states, thus facilitating a comparison of the calculated and
experimental spectra in a wide energy region.

Valence band photoemission spectroscopy (PES) using high photon
energies and O $1s \rightarrow 2p$ x-ray absorption spectroscopy
(XAS) were performed on the beamline BL25SU at the SPring-8
synchrotron radiation facility. The PES spectra were taken using a GAMMADATA-SCIENTA
SES-200 electron energy analyzer and XAS spectra were measured by
the total electron yield. The overall energy resolution
was set to 0.2 eV.  The pressure in the analyzer chamber was about
4 x 10$^{-8}$ Pa. Single crystals of SrVO$_3$ were cooled to 20 K
and clean surfaces were obtained by fracturing {\it in situ}
for the PES spectra, and by scraping {\it in situ} for the much
more bulk sensitive XAS spectra. Well-oriented single-crystalline
V$_2$O$_3$ samples were cleaved {\it in situ} at a temperature
near the metal-insulator transition, yielding clean
specular surfaces.  The surface cleanliness was confirmed before
and after the spectral run.

\begin{figure}
\centering
\includegraphics[clip=true,width=0.5\textwidth,angle=270]{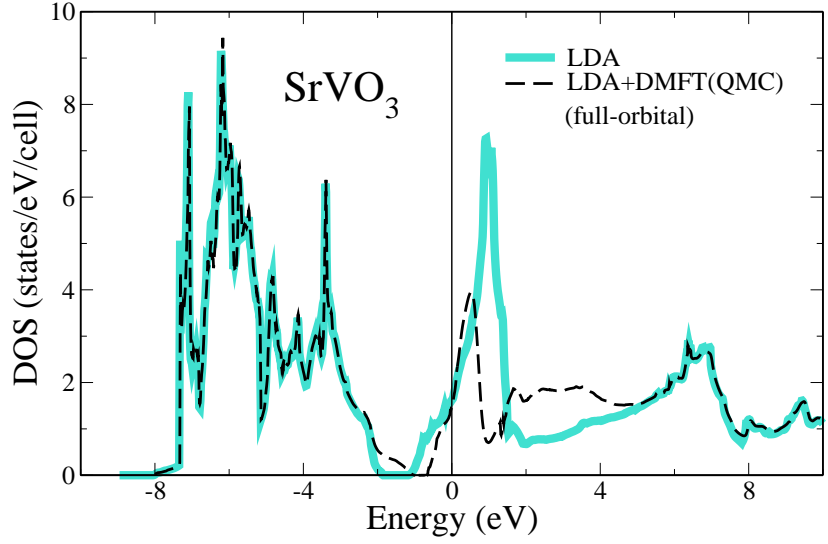}
\caption{Comparison of total spectral functions of SrVO$_3$ calculated via:
LDA - full light line, using the full-orbital self-energy from LDA+DMFT(QMC) -
dashed black line. Fermi level corresponds to zero.}
\label{SrVO3_total}
\end{figure}

\begin{figure}
\centering
\includegraphics[clip=true,width=0.5\textwidth,angle=270]{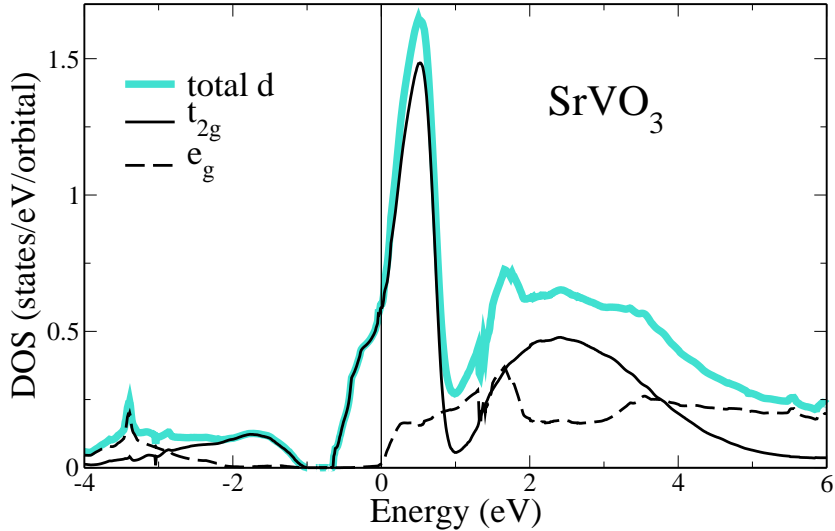}
\caption{Partial V-3$d$ spectral functions for SrVO$_3$ calculated
using the full-orbital self-energy from LDA+DMFT(QMC). Total $d$ - full light line,
t$_{2g}$ - full black line, e$_g$ - dashed black line. Fermi level corresponds
to zero.}
\label{SrVO3_d}
\end{figure}

We start with the results for the SrVO$_3$ system. In Fig.~\ref{SrVO3_sigma},
the self-energy on the real energy axis for SrVO$_3$ calculated
via~(\ref{acSE}) is shown. The $m^*/m$ ratio is calculated via $m^*/m =
1/(1-(\partial\Re\Sigma(\omega)/\partial\omega))$. This self-energy was used
for the calculation of total and partial DOSs in the full-orbital Hilbert space
(see Fig.~\ref{SrVO3_total},\ref{SrVO3_d}).
%%%%%%%%%%%%%%%%%%%%%%%%%%%%%%%%%%%%%%%%%%%%%%%%%%%%%%%%%%%%%%%%%%%%%%%%%%%%%
%%%%%%%%%%%%%%%%%%%%%%%%%%%%%%%%%%%%%%%%%%%%%%%%%%%%%%%%%%%%%%%%%%%%%%%%%%%%%
%%%%%%%%%%%%%%%%%%%%%%%%%%%%%%%%%%%%%%%%%%%%%%%%%%%%%%%%%%%%%%%%%%%%%%%%%%%%%
%%%%%%%%%%%%%%%%%%%%%%%%%%%%%%%%%%%%%%%%%%%%%%%%%%%%%%%%%%%%%%%%%%%%%%%%%%%%%
In Fig.~\ref{SrVO3_total}, the total spectral functions of SrVO$_3$ are
presented. Differences between LDA and
full-orbital LDA+DMFT(QMC) spectral functions
mainly occur near the Fermi level. In particular, the LDA spectral function has a more
pronounced quasiparticle peak. The DOS calculated using the
full-orbital self-energy has a three-peak structure: lower
Hubbard band (suppressed by oxygen states), quasiparticle peak and
upper Hubbard band located at about 3 eV. The origin of the upper
Hubbard band becomes clear from Fig.~\ref{SrVO3_d}. One can see that this
broad peak is the sum of V-3$d$(e$_g^\sigma$) and V-3$d$(t$_{2g}$) states.
It should be noted that correlations on
the V-3$d$(e$_g^\sigma$) orbitals were not explicitly included here.
We find, however, that the V-3$d$(e$_g^\sigma$) states are slightly
modified by the full-orbital
self-energy due to the of hybridization with the correlated V-3$d$(t$_{2g}$)
orbitals. The question how correlations affect the position and width of
V-3$d$(e$_g^\sigma$) states directly can only be answered by employing the
full 3$d$-shell 5x5 projected Hamiltonian.

Introducing correlations between t$_{2g}$ states changes
significantly the total and partial LDA DOSs of SrVO$_3$. The main
modification is a transfer of spectral weight from the energy region
near Fermi level to the lower and upper Hubbard bands, and the reduction
of the weight of the quasiparticle peak. Our calculations find
a strongly correlated but still metallic ground state for SrVO$_3$.

%%%%%%%%%%%%%%%%%%%%%%%%%%%%%%%%%%%%%%%%%%%%%%%%%%%%%%%%%%%%%%%%%%%%%%%%%%%%%
\begin{figure}
\centering
\includegraphics[clip=true,width=0.5\textwidth,angle=270]{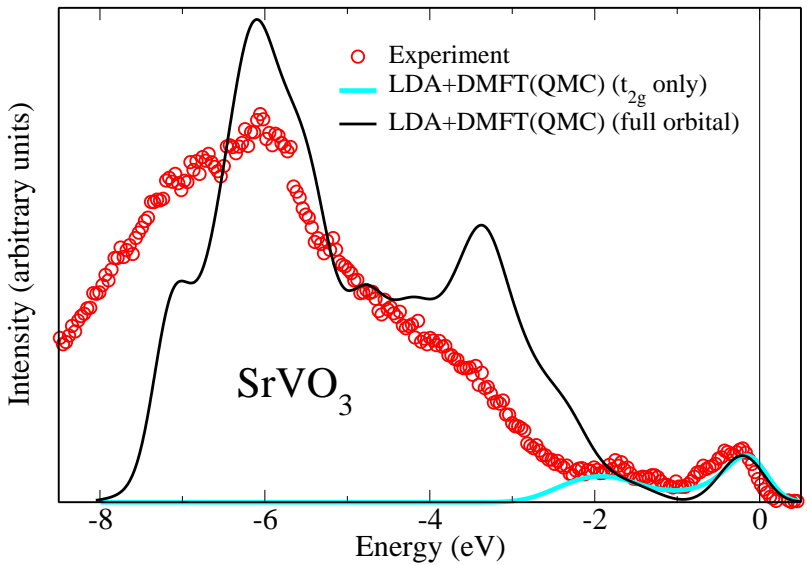}
\caption{Comparison of photoemission spectra of SrVO$_3$ with spectral functions
calculated via LDA+DMFT(QMC): taking into account only t$_{2g}$ - light line,
using full-orbital self-energy - black line. Theoretical spectra are multplied with
the Fermi function corresponding to 20K and broadened with a 0.2 eV Gaussian to
account for the instrumental resolution. Intensities are normalized for quasiparticle
peaks. Fermi level corresponds to zero.}
\label{compar_PES_SrVO3}
\end{figure}
%%%%%%%%%%%%%%%%%%%%%%%%%%%%%%%%%%%%%%%%%%%%%%%%%%%%%%%%%%%%%%%%%%%%%%%%%%%%%

To compare our results with the experimental PES we
calculated a weighted sum of V-3$d$ and O-2$p$ spectral functions according
to the photoemission cross section ratio 3:1, corresponding to an
experimental photon energy 900~eV. The theoretical spectra were
multiplied with the Fermi function corresponding to 20 K and broadened
with a 0.2 eV Gaussian to take into account the instrumental resolution. In
Fig.~\ref{compar_PES_SrVO3} one can see that the full-orbital
spectra obtained in this way describe not only the quasiparticle peak,
but also the peak at
-6~eV and the shoulder at -3.5~eV of the PES spectra. Since previous
LDA+DMFT(QMC) results were taking into account only the t$_{2g}$ states
they were not able to describe PES spectra below -2~eV. It is interesting
to note that previous experimental studies did not find any states at
-3.5~eV; only the new spectra reported here show this feature.

%%%%%%%%%%%%%%%%%%%%%%%%%%%%%%%%%%%%%%%%%%%%%%%%%%%%%%%%%%%%%%%%%%%%%%%%%%%%%
\begin{figure}
\centering
\includegraphics[clip=true,width=0.5\textwidth,angle=0]{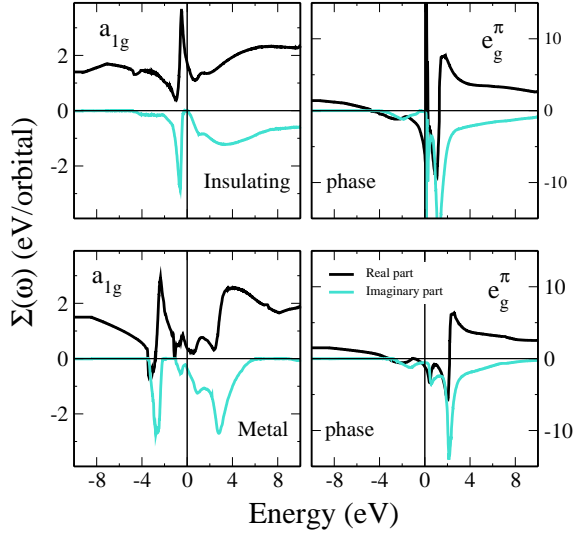}
\caption{Self-energy on real energy axis for V$_2$O$_3$. Real part of the
self-energy from LDA+DMFT(QMC) - full black line, imaginary part - full
light line. Fermi level corresponds to zero.}
\label{V2O3_sigma}
\end{figure}
%%%%%%%%%%%%%%%%%%%%%%%%%%%%%%%%%%%%%%%%%%%%%%%%%%%%%%%%%%%%%%%%%%%%%%%%%%%%%

The influence of correlation effects on the electronic structure will be more
pronounced for systems close to the Mott insulator transition. For this
purpose we compare the total and partial DOSes for V$_2$O$_3$ in the insulating
and metallic phases calculated via LDA and full-orbital LDA+DMFT \cite{U-value}.
Since the V-3$d$(t$_{2g}$) orbitals in trigonal V$_2$O$_3$ split into
non-equivalent a$_{1g}$ and e$_{g}^{\pi}$ states, the self-energy will be
different for these orbitals (see fig~\ref{V2O3_sigma}).
%%%%%%%%%%%%%%%%%%%%%%%%%%%%%%%%%%%%%%%%%%%%%%%%%%%%%%%%%%%%%%%%%%%%%%%%%%%%%
\begin{figure}
\centering
\includegraphics[clip=true,width=0.5\textwidth,angle=0]{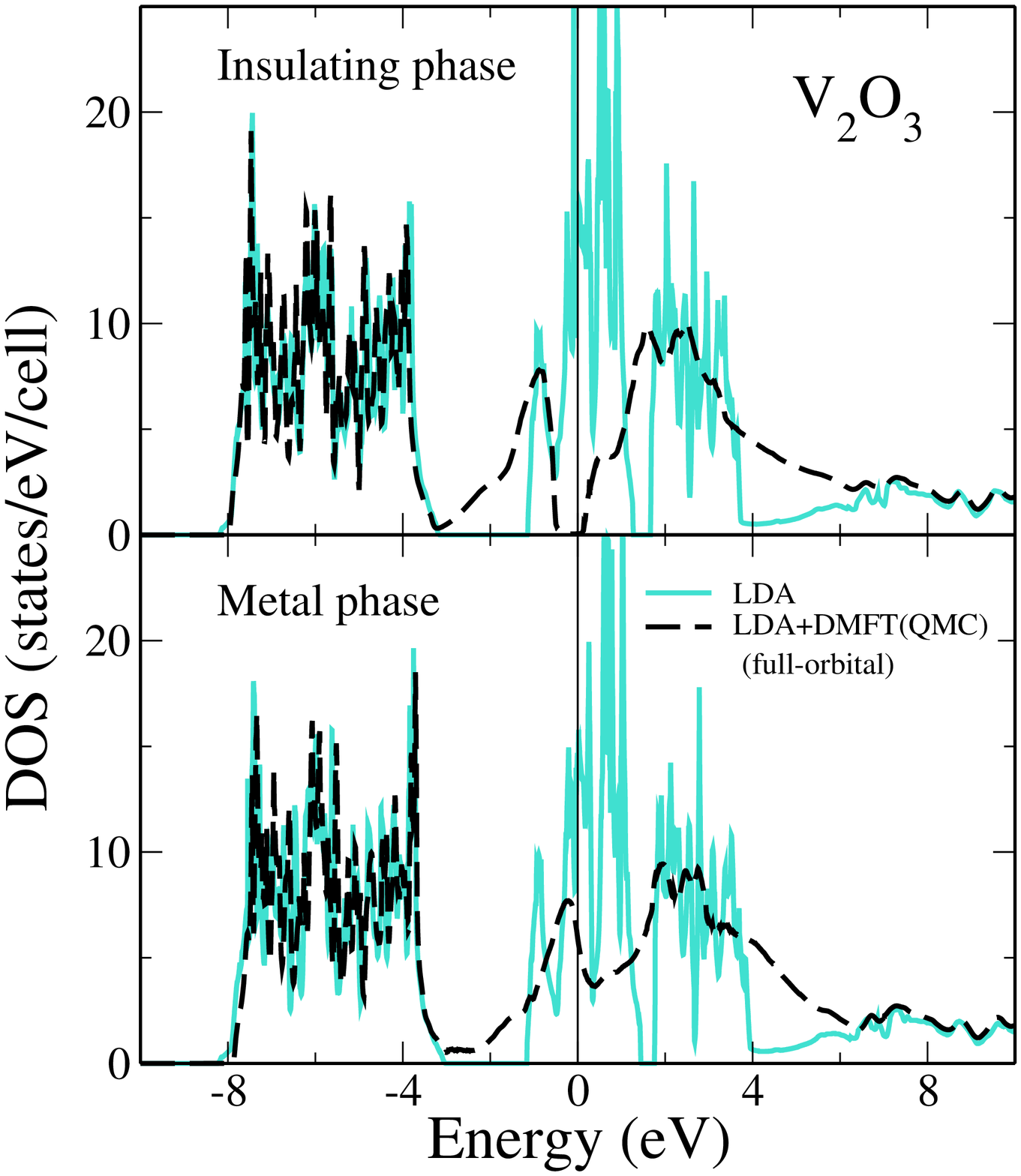}
\caption{Comparison of total spectral functions of V$_2$O$_3$ calculated via:
LDA - full light line, using full-orbital self-energy from LDA+DMFT(QMC)-
dashed black line. Fermi level corresponds to zero.}
\label{V2O3_total}
\end{figure}
%%%%%%%%%%%%%%%%%%%%%%%%%%%%%%%%%%%%%%%%%%%%%%%%%%%%%%%%%%%%%%%%%%%%%%%%%%%%%
%%%%%%%%%%%%%%%%%%%%%%%%%%%%%%%%%%%%%%%%%%%%%%%%%%%%%%%%%%%%%%%%%%%%%%%%%%%%%
\begin{figure}
\centering
\includegraphics[clip=true,width=0.5\textwidth,angle=0]{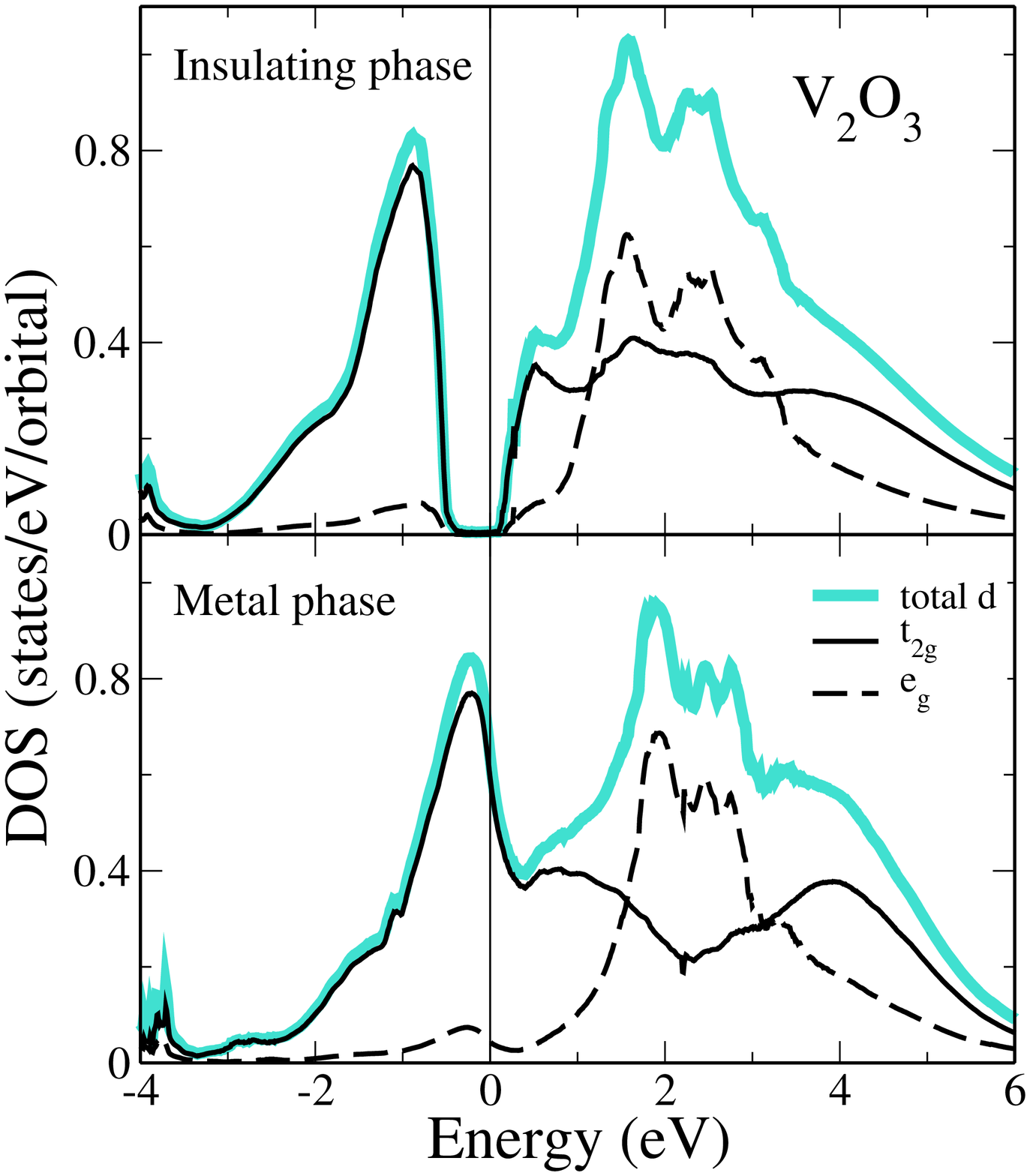}
\caption{Partial V-3$d$ spectral functions for V$_2$O$_3$ calculated using
full-orbital self-energy from LDA+DMFT(QMC). Total d - full light line,
t$_{2g}$ - full black line, e$_g$ - dashed black line. Fermi level
corresponds to zero.}
\label{V2O3_d}
\end{figure}
%%%%%%%%%%%%%%%%%%%%%%%%%%%%%%%%%%%%%%%%%%%%%%%%%%%%%%%%%%%%%%%%%%%%%%%%%%%%%
In Fig.~\ref{V2O3_total} one can see that the introduction of correlation effects
changes the total DOS drastically. Whereas the LDA DOS is metallic
for both metallic and insulating phases, the LDA+DMFT spectra
clearly show the metal-insulator transition. Another important feature are the
modifications near the Fermi energy and in the energy
window from -3~eV to 7~eV. Moreover, there are also changes in the
e$_g^\sigma$ states which are caused by the hybridization with the
correlated t$_{2g}$ bands, and are not due to a direct calculation of
correlations in the e$_g^\sigma$ states. In Fig.~\ref{V2O3_d}
the total $d$-state DOS above the Fermi energy, consisting of t$_{2g}$ and
e$_g^\sigma$ states is shown.
%%%%%%%%%%%%%%%%%%%%%%%%%%%%%%%%%%%%%%%%%%%%%%%%%%%%%%%%%%%%%%%%%%%%%%%%%%%%%
\begin{figure}
\centering
\includegraphics[clip=true,width=0.5\textwidth,angle=0]{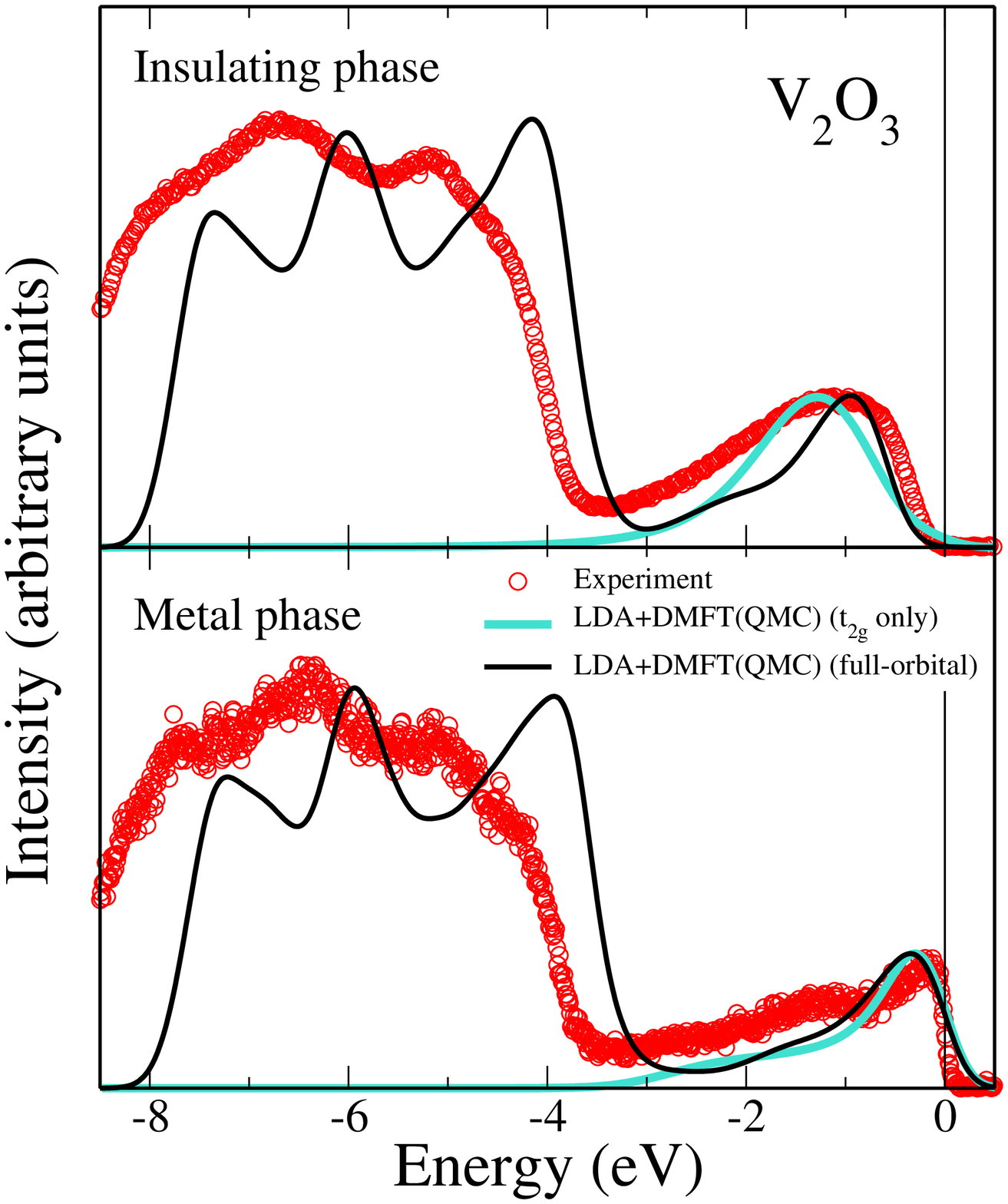}
\caption{Comparison of photoemission spectra of V$_2$O$_3$ with
spectral functions calculated via LDA+DMFT(QMC): taking into account only
t$_{2g}$ - light line, using full-orbital self-energy - black line. Theoretical
spectra are multiplied with the Fermi function and broadened with a 0.2
eV Gaussian to simulate instrumental resolution. Intensities are normalized
for peaks situated around -1~eV. Fermi level corresponds to zero.}
\label{compar_PES_V2O3}
\end{figure}
%%%%%%%%%%%%%%%%%%%%%%%%%%%%%%%%%%%%%%%%%%%%%%%%%%%%%%%%%%%%%%%%%%%%%%%%%%%%%

In Fig.~\ref{compar_PES_V2O3} a comparison of experimental
PES and calculated LDA+DMFT(QMC) (t$_{2g}$ only and
full-orbital) spectra is presented. The full-orbital spectrum is
a weighted sum of the calculated V-3$d$ and O-2$p$ spectral functions,
according to the photoemission cross section ratio 2:1,
corresponding to the experimental photon energy of 500~eV. The
theoretical spectra were multiplied with the Fermi function and
broadened with a Gaussian of 0.2 eV.\cite{Allen04}
Below the Fermi energy the LDA+DMFT(QMC) spectral functions
(for t$_{2g}$ only and the full-orbital scheme) agree quite well with the PES
spectra. However, the theory curves do not yet describe fine details.
Specifically, the PES spectrum shows two definite slope changes, at roughly
-0.6~eV and -1.5~eV, producing a rather flat topped spectrum
centered on -1~eV, whereas the theory curves show a single peak
centered on -1~eV.

%%%%%%%%%%%%%%%%%%%%%%%%%%%%%%%%%%%%%%%%%%%%%%%%%%%%%%%%%%%%%%%%%%%%%%%%%%%%%
\begin{figure}
\centering
\includegraphics[clip=true,width=0.5\textwidth,angle=0]{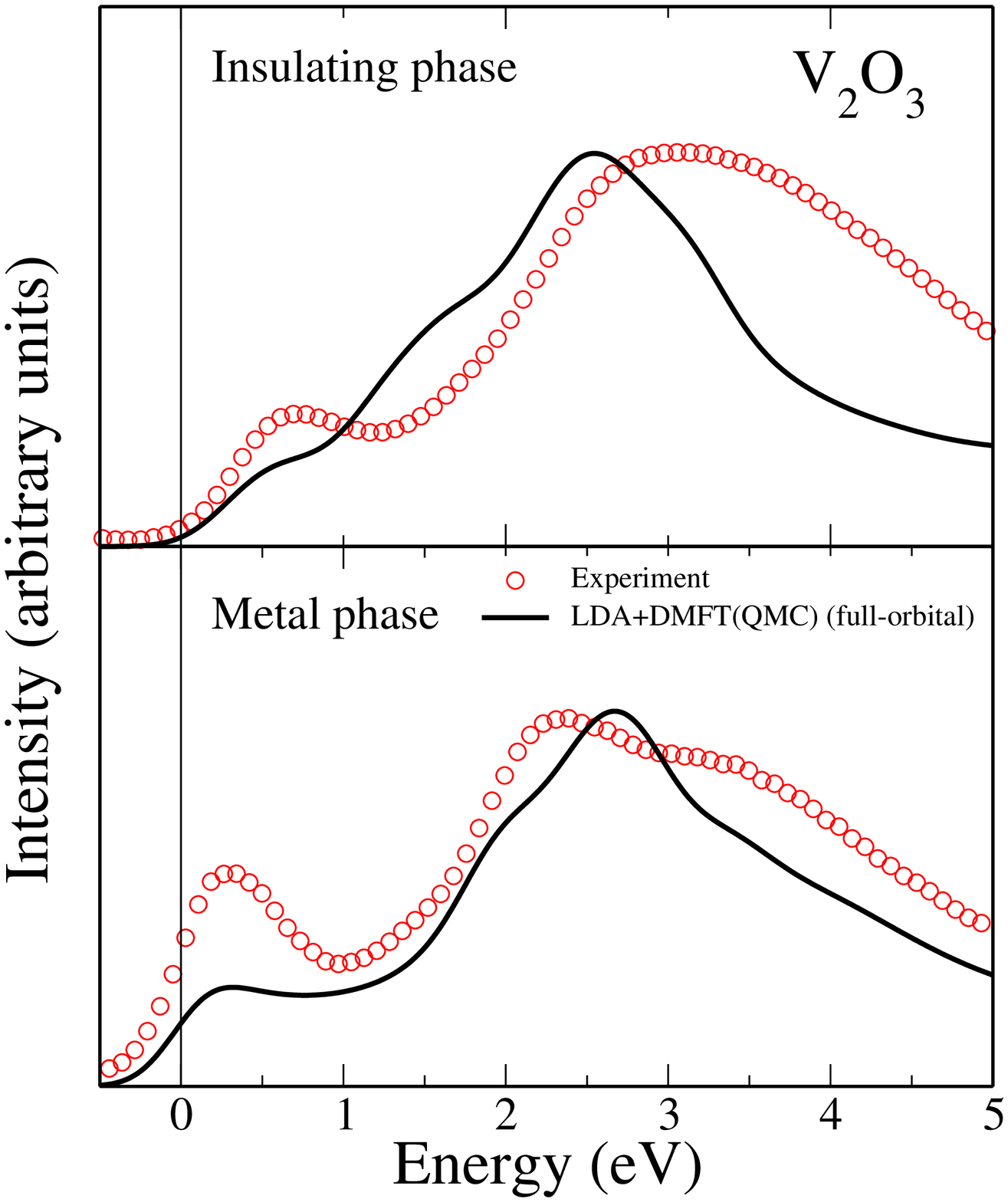}
\caption{Comparison of O $1s \rightarrow 2p$ x-ray absorption spectroscopy (XAS)
spectra of V$_2$O$_3$ with spectral functions calculated via LDA+DMFT(QMC)
%%%%: taking into account only t$_{2g}$ - light line,
using full-orbital self-energy.
%%%%%- black line.
Theoretical spectra are multiplied with the Fermi function and
broadened with a 0.2 eV Gaussian to simulate instrumental resolution.
Intensities of the theory curves are normalized to the correct number of
V-3$d$ electrons in the unit cell
%%%%4 electrons for the t$_{2g}$,
(8 electrons).
%%%%%for the full orbital spectrum), the
The experimental curve is normalized to the same peak height as
the full orbital curve. Fermi level corresponds to zero.}
\label{compar_XAS_V2O3}
\end{figure}
%%%%%%%%%%%%%%%%%%%%%%%%%%%%%%%%%%%%%%%%%%%%%%%%%%%%%%%%%%%%%%%%%%%%%%%%%%%%%

A comparison of the calculated LDA+DMFT(QMC) full-orbital spectral functions
and the O $1s \rightarrow 2p$ XAS spectrum is shown on
Fig.~\ref{compar_XAS_V2O3}. The full-orbital spectra (which are the
partial O-2$p$ spectra for the unoccupied V-3$d$ states) are found to be in
rather good agreement with the structures of the experimental spectrum
their relative intensities and their energies. This is due to the inclusion
of O-2$p$ and V-3$d$($e_g^\sigma$) states in the calculations.  The
strong hybridization of the O-2$p$ with the d-states is described
more correctly in the full-orbital calculations and the inclusion
of $e_g^\sigma$ states (even without correlations)
significantly changes the V-3$d$-shell in the energy
regions above the Fermi edge. The inclusion of correlations in
$e_g^\sigma$ states within a 5x5 projected V-3$d$ Hamiltonian is
expected to add spectral weight in the energy interval between 3 eV and 5 eV
(see Fig.~\ref{compar_XAS_V2O3}).

%%%%%%%%%%%%%%%%%%%%%%%%%%%%%%%%%%%%%%%%%%%%%%%%%%%%%%%%%%%%%%%%%%%%%%%%%%%%
% Conclusion
%%%%%%%%%%%%%%%%%%%%%%%%%%%%%%%%%%%%%%%%%%%%%%%%%%%%%%%%%%%%%%%%%%%%%%%%%%%%
\section{Conclusion}
\label{conclusion}

We formulated a fully {\it{ab initio}} and self-consistent
computational scheme based on Wannier functions (WF) for the
calculation of the electronic structure of strongly correlated
compounds. The WF formalism provides an explicit strategy for the
construction of the matrix elements of the required operators
(Hamiltonian, self-energy operator, etc.), both in full-orbital
and few-orbital bases, in real and reciprocal representations. The
WF formalism allows one to project these operators from the
full-orbital space to the few-orbital space and back, keeping the
complete information about all orbitals. These projections are the
essential ingredients of the computational scheme presented here.
The self-consistency involves not only the self-energy but also
the WF basis itself, the charge density with the LDA-potential and the
interaction strength parameters between the electrons on the WFs.
The spectra obtained thereby found to be in good agreement with
new bulk-sensitive experimental data.

In the present work we did not yet employ the full scheme (see
Fig.~\ref{scheme}) to investigate spectral functions of strongly
correlated systems but employed only few-orbital Hamiltonians with
t$_{2g}$ symmetry. Clearly, full  $d$-shell DMFT(QMC) results
including e$_g$ states will provide additional information about
correlation effects in the system. Such studies, as well as
constrained LDA calculations in the WF basis and investigations of
the feedback from DMFT to LDA part are in progress now.

%%%%%%%%%%%%%%%%%%%%%%%%%%%%%%%%%%%%%%%%%%%%%%%%%%%%%%%%%%%%%%%%%%%%%%%%%%%%
% Acknowledgement
%%%%%%%%%%%%%%%%%%%%%%%%%%%%%%%%%%%%%%%%%%%%%%%%%%%%%%%%%%%%%%%%%%%%%%%%%%%%
\section{Acknowledgement}

This work was supported by Russian Basic Research Foundation grant
RFFI-GFEN-03-02-39024\_a, RFFI-04-02-16096 and by the Deutsche Forschungsgemeinschaft
through Sonderforschungsbereich 484,
Augsburg, the John von Neumann Institut f\"ur Computing, J\"ulich,  and in
part by the joint UrO-SO project N22, Grant of
President of Russian Federation for young scientists MK-95.2003.02, Dynasty
Foundation and International Center for Fundamental Physics in Moscow program
for young scientists 2004), Russian Science Support Foundation program for
young PhD of Russian Academy of Science 2004. Work at the University of
Michigan was supported by the U.S. NSF under (Grant No.~DMR-03-02825) and
at POSTECH by KOSEF through eSSC.
All soft X-ray absorption and photoemission studies have been performed in SPring-8
under the upproval and support of Japan Syncrotron Radiation Research Institute.
\appendix

%%%%%%%%%%%%%%%%%%%%%%%%%%%%%%%%%%%%%%%%%%%%%%%%%%%%%%%%%%%%%%%%%%%%%%%%%%%%
% Hilbert transformation
%%%%%%%%%%%%%%%%%%%%%%%%%%%%%%%%%%%%%%%%%%%%%%%%%%%%%%%%%%%%%%%%%%%%%%%%%%%%
\section{Hilbert transformation}
\label{hilbert}

For cubic systems the matrix of the self-energy operator (for example for
$t_{2g}$ orbitals) is diagonal and all diagonal elements are equal.
Therefore the calculation of the Green function within DMFT by
integration of the Hamiltonian over the BZ is equivalent to the Hilbert
transformation of the non-interacting (LDA) DOS $N^0(\epsilon)$:
%%%%%%%%%%%%%%%%%%%%%%%%%%%
\begin{eqnarray}
\Sigma(\omega) = \left( \begin{array}{cccc} \sigma(\omega) & 0 &
\ldots & 0 \\ 0 & \sigma(\omega) & \ldots & 0 \\ \vdots & \vdots &
\ddots & \vdots \\ 0 & 0 & \ldots & \sigma(\omega)
\end{array} \right) \Rightarrow
G(\omega)=\int\limits_{IBZ}(\omega - \Sigma(\omega) -
H^0_{LDA}(k))^{-1}dk = \\ \nonumber = G^0(\omega - \sigma(\omega))
= \int \frac{N^0(\epsilon)}{\omega - \sigma(\omega) -
\epsilon}d\epsilon. \nonumber
\end{eqnarray}
%%%%%%%%%%%%%%%%%%%%%%%%%%%

%%%%%%%%%%%%%%%%%%%%%%%%%%%%%%%%%%%%%%%%%%%%%%%%%%%%%%%%%%%%%%%%%%%%%%%%%%%%
% Self-energy on the real energy axis
%%%%%%%%%%%%%%%%%%%%%%%%%%%%%%%%%%%%%%%%%%%%%%%%%%%%%%%%%%%%%%%%%%%%%%%%%%%%
\section{Self-energy on the real energy axis}
\label{continuation}

DMFT produces Green functions and self-energies on the imaginary energy axis.
With the maximum entropy method \cite{MEM}, the spectral function on the real
energy axis is calculated, which yields the imaginary part of the Green function.
In order to obtain the self-energy on the real energy axis, the full complex
Green function $G(\varepsilon)$ is calculated using its imaginary part obtained
by MEM:
%%%%%%%%%%%%%%%%%%%%%%%%%%%
\begin{eqnarray}
\label{fullGF} G(\varepsilon) =
-\frac{1}{\pi}\int\limits_{-\infty}^{\infty} \frac{\Im
G(\varepsilon')d\varepsilon '}{\varepsilon - \varepsilon' + i\eta}.
\end{eqnarray}
%%%%%%%%%%%%%%%%%%%%%%%%%%%
The self-energy for real energies is then calculated by solving the following
two equations with the two variables $\Re \Sigma(\varepsilon)$ and
$\Im \Sigma(\varepsilon)$ :
%%%%%%%%%%%%%%%%%%%%%%%%%%%
\begin{eqnarray}
\label{acSE} \Re ,\Im\{G(\varepsilon)\} = \Re
,\Im\{\int\limits_{BZ} (\varepsilon - H({\bf k}) -
\Sigma(\varepsilon))^{-1}d{\bf k}\},
\end{eqnarray}
where
\begin{eqnarray}
\Sigma(\varepsilon) = \Re \Sigma(\varepsilon) + iIm
\Sigma(\varepsilon).
\end{eqnarray}
%%%%%%%%%%%%%%%%%%%%%%%%%%%

%%%%%%%%%%%%%%%%%%%%%%%%%%%%%%%%%%%%%%%%%%%%%%%%%%%%%%%%%%%%%%%%%%%%%%%%%%%%
% Non-orthogonal basis set
%%%%%%%%%%%%%%%%%%%%%%%%%%%%%%%%%%%%%%%%%%%%%%%%%%%%%%%%%%%%%%%%%%%%%%%%%%%%
\section{Non-orthogonal basis set}
\label{ortho}

Eq. (\ref{WF}) is valid only for orthogonal LMTO orbitals. In the
case of general non-orthogonal LMTOs (or any other atomic type orbital basis
set), an orthogonalization procedure can be used by defining an orthogonal
Hamiltonian $\widetilde H$ and the corresponding eigenvectors $\widetilde C$
for the non-orthogonal Hamiltonian $H$ and overlapping matrix $O$:
%%%%%%%%%%%%%%%%%%%%%%%%%%%
\begin{eqnarray}
\label{H_ort} \widetilde H = O^{-\frac{1}{2}}H O^{-\frac{1}{2}},
\\ \nonumber \widetilde C = O^{\frac{1}{2}} C.
\end{eqnarray}
%%%%%%%%%%%%%%%%%%%%%%%%%%%
This orthogonalization is equivalent to the basis set transformation:
%%%%%%%%%%%%%%%%%%%%%%%%%%%
\begin{eqnarray}
\label{phi_ort} |\widetilde\phi_n\rangle  =
\sum_{n'}O^{-\frac{1}{2}}_{nn'}|\phi_{n'}\rangle.
\end{eqnarray}
%%%%%%%%%%%%%%%%%%%%%%%%%%%
Then for the non-orthogonal LMTO the trial function $|\phi_n\rangle$ has to
be replaced by $|\widetilde\phi_n\rangle$ and the eigenvectors with coefficients
$c^{\bf k}_{ji}$ by $\widetilde c^{\bf k}_{ji}$.

%%%%%%%%%%%%%%%%%%%%%%%%%%%%%%%%%%%%%%%%%%%%%%%%%%%%%%%%%%%%%%%%%%%%%%%%%%%%
% thebibliography
%%%%%%%%%%%%%%%%%%%%%%%%%%%%%%%%%%%%%%%%%%%%%%%%%%%%%%%%%%%%%%%%%%%%%%%%%%%%


\begin{thebibliography}{99}
\bibitem{efermi} In the calculations performed in the present paper
we assumed the position of the Fermi level to be unchanged when
going from LDA to DMFT(QMC). It was found in previous
investigations \cite{LDADMFT} that this is a good approximation
for vanadates.
\bibitem{nearly} Or nearly diagonal, if the symmetry is lowered by a
relatively small distortion of the crystal structure.
\bibitem{U-value} For illustration purposes we chose the most insulating
(U = 5.5 eV) for insulating V$_2$O$_3$ and the most metallic solutions
(U = 4.5) for metallic V$_2$O$_3$.
\end{thebibliography}
\end{document}